\documentclass[aps,pre,floatfix,reprint,groupedaddress,showkeys,superscriptaddress]{revtex4-2}
\usepackage{graphicx}
\usepackage{dcolumn}
\usepackage{bm}
\usepackage{epstopdf}
\usepackage{amsmath}
\usepackage{float}
   \usepackage[all]{xy}
\usepackage{subfigure}
\usepackage{xcolor}

\newcommand{\beginsupplement}{%
	\setcounter{table}{0}
	\renewcommand{\thetable}{S\arabic{table}}%
	\setcounter{figure}{0}
	\renewcommand{\thefigure}{S\arabic{figure}}%
	\setcounter{equation}{0}
	\renewcommand{\theequation}{S\arabic{equation}}
}

\begin{document}
	
\title{Toward a universal model for spatially structured populations}

 \author{Loïc Marrec}
 \altaffiliation[Present address: ]{Institute of Ecology and Evolution, University of Bern, Baltzerstrasse 6, CH-3012 Bern, Switzerland}
   \affiliation{Sorbonne Universit{\'e}, CNRS, Institut de Biologie Paris-Seine, Laboratoire Jean Perrin (UMR 8237), F-75005 Paris, France}
   
 \author{Irene Lamberti}
   \altaffiliation[Present address: ]{Institute of Bioengineering, School of Life Sciences, École Polytechnique Fédérale de Lausanne (EPFL), CH-1015 Lausanne, Switzerland}
   \affiliation{Sorbonne Universit{\'e}, CNRS, Institut de Biologie Paris-Seine, Laboratoire Jean Perrin (UMR 8237), F-75005 Paris, France}
 
 \author{Anne-Florence Bitbol}
   \email[Corresponding author: ]{anne-florence.bitbol@epfl.ch}
 \affiliation{Sorbonne Universit{\'e}, CNRS, Institut de Biologie Paris-Seine, Laboratoire Jean Perrin (UMR 8237), F-75005 Paris, France}
 \affiliation{Institute of Bioengineering, School of Life Sciences, École Polytechnique Fédérale de Lausanne (EPFL), CH-1015 Lausanne, Switzerland}
  \affiliation{SIB Swiss Institute of Bioinformatics, CH-1015 Lausanne, Switzerland}

\begin{abstract}
A key question in evolution is how likely a mutant is to take over. This depends on natural selection and on stochastic fluctuations. Population spatial structure can impact mutant fixation probabilities. We introduce a model for structured populations on graphs that generalizes previous ones by making migrations independent of birth and death. We demonstrate that by tuning migration asymmetry, the star graph transitions from amplifying to suppressing natural selection. The results from our model are universal in the sense that they do not hinge on a modeling choice of microscopic dynamics or update rules. Instead, they depend on migration asymmetry, which can be experimentally tuned and measured.
\end{abstract}

\maketitle

\emph{Introduction.---} Classical models of well-mixed, homogeneous microbial populations assume that each microorganism competes with all others. However, this simplification holds in few natural situations. For instance, during an infection, microbial populations are subdivided between different organs~\cite{VanMarle07,Schnell10} and hosts. Any spatial structure, e.g. that of a Petri dish, implies a stronger competition between neighbors than between distant individuals. Even well-agitated liquid suspensions feature deviations compared to idealized well-mixed populations~\cite{Herrerias18}.

Spatial structure can have major consequences on evolution.  Remarkably, the fixation probability of a mutant can be affected, with specific structures amplifying or suppressing natural selection~\cite{Lieberman05}. Studying these effects requires going beyond simple structures~\cite{Wright31,Kimura64} where migration is symmetric between demes (i.e. subpopulations), since fixation probabilities are unaffected in these cases~\cite{Maruyama70,Maruyama74,Slatkin81}, unless extinctions of demes occur~\cite{Barton93}. Ref.~\cite{Lieberman05} introduced a seminal model for complex structures, known as evolutionary dynamics on graphs, with one individual at each node of a graph, and probabilities that their offspring replaces a neighbor along each edge of the graph. However, in such models, evolutionary outcomes can drastically depend on the details of the microscopic dynamics or ``update rule'', e.g. whether the individual that divides or the one that dies is chosen first, even if selection always acts at division~\cite{Antal06, Kaveh15, Hindersin15, Pattni15}. This lack of universality raises issues for applicability to real populations, where one birth does not necessarily entail one death and vice-versa. Furthermore, in most microbial populations, individuals freely compete with their closest neighbors, motivating a coarse-grained description, with demes rather than individuals on graph nodes~\cite{Wright31,Kimura64,Campos06,Houchmandzadeh11,Houchmandzadeh13,Constable14}. Current experiments with well-mixed demes at each node of a star graph~\cite{Kassen} require theoretical predictions with realistic microscopic dynamics.

We propose a model for complex spatial population structures where migrations are independent from birth and death events. We investigate the fixation probability of mutants in the rare migration regime. We demonstrate that migration asymmetry determines whether the star graph amplifies or suppresses natural selection. We find a mapping to the model of Ref.~\cite{Lieberman05} under specific constraints on migration rates. 

\emph{Model.---} We model a structured population as a directed graph where each node $i\in\{1,\dots,D\}$ contains a well-mixed deme with carrying capacity $K$, and migration rates $m_{ij}$ per individual from deme $i$ to deme $j\neq i$ are specified along each edge $ij$. We then address populations including demes with different carrying capacities~\cite{Supplement}. We consider microorganisms with two types, wild-type (W) and mutant (M), with fitnesses and death rates denoted by $f_a$ and $g_a$, where $a=W$ or $a=M$. Here, we call fitness the maximal division rate of microorganisms, reached in exponential growth. Their division rate in deme $i$ is given by the logistic function $f_a(1-N_i/K)$, where $N_i$ is the number of individuals in deme $i$. We take wild-type fitness as a reference, $f_W=1$. We address selection on birth, and hence $g_M=g_W$, but our results can be generalized to selection on death. We focus on the regime where deme sizes $N_i$ fluctuate weakly around their deterministic steady-state values, without extinctions~\cite{Barton93,Whitlock97,Whitlock03}.

We assume that mutations are rare enough for further mutation events to be neglected while the fate of a given mutant lineage (taking over or disappearing) is determined. We consider an initial mutant placed uniformly at random, which is realistic for spontaneous mutations occurring either with a fixed rate or with a fixed probability upon division. Note that in models with one individual per node, uniform initialization is more appropriate in the first case, while placing mutants proportionally to the replacement probability of a node (``temperature initialization'') is more appropriate in the second one~\cite{Adlam15}. This distinction vanishes here, as division rate does not depend on location. Under uniform initialization, the fixation probability of a neutral mutant is independent of structure for connected graphs~\cite{Supplement}. Compared to the well-mixed population with the same total size, an \textit{amplifier} of natural selection features a larger fixation probability for beneficial mutants ($f_M>f_W$), and a smaller one for deleterious mutants ($f_M<f_W$), while a \textit{suppressor} has the opposite characteristics~\cite{Allen20}. 

We focus on the rare migration regime~\cite{Slatkin81}, where fixation of a type (W or M) in a deme is much faster than migration timescales. Then, the state of the population can be described in a coarse-grained way by whether each deme is mutant or wild-type. Its evolution is a Markov process where elementary steps are migration events, which change the state of the system if fixation ensues. Then, a mutant first needs to fix in the deme where it appeared, before mutants can spread to other demes. Since fixation in a homogeneous deme is well-known, we study the second stage, starting from one fully mutant deme.

\emph{Link with models with one individual per node.---} A~formal mapping can be made between our model and that of~\cite{Lieberman05}, if the same graph is considered, with a deme per node in our model and with one individual per node in~\cite{Lieberman05} (see~\cite{Supplement}). The probability $\mathcal{P}_{i\rightarrow j}$ that, upon a migration event resulting into fixation, an individual from deme $i$ takes over in deme $j$ in our model maps to the probability $P^{[4]}_{i\rightarrow j}$ that, upon a division, the offspring from node $i$ replaces the individual on node $j$ in the model of~\cite{Lieberman05}: 
\begin{equation} \label{mapf}
\mathcal{P}_{i\rightarrow j} = \frac{m_{ij}N_i\rho_{i}}{\sum_{k,\,l}m_{kl}N_k\rho_{k}} \,\,\,\,\,\leftrightarrow \,\,\,\,\,P^{[4]}_{i\rightarrow j} = \frac{w_{ij}f_i}{\sum_{k,\,l} w_{kl}f_k}\,.
\end{equation}
In this mapping, the product $N_i\rho_i$ of deme size $N_i$ and fixation probability $\rho_i$ of an organism from deme $i$ in our model plays the part of fitness $f_i$ of the individual on node $i$ in~\cite{Lieberman05}, while the migration rate $m_{ij}$ plays the part of the replacement probability $w_{ij}$ that the offspring of the individual in $i$ replaces that in $j$. However, an important constraint in the ``Birth-death'' model of~\cite{Lieberman05} (also known as biased invasion process~\cite{Antal06}) is $\sum_j w_{ij}=1$ for all $i$, because replacement includes birth, migration and death at once, and population size is constant. By contrast, migration rates $m_{ij}$ in our model are all independent.

A generalized circulation theorem holds for our model~\cite{Supplement}, in the spirit of~\cite{Lieberman05}. Specifically, a population of $D$ demes on a graph has the same mutant fixation probability as the clique if and only if, for all nodes of the graph, the total outgoing  migration rate is equal to the total incoming migration rate. 

Thus, we expect fixation probabilities in our model to map to those of~\cite{Lieberman05} for circulations or if $\sum_j m_{ij}$ is independent of $i$, but to potentially differ otherwise. We now consider specific graphs with strong symmetries. 

\begin{figure}[htb]
	\centering
	\includegraphics[width=\columnwidth]{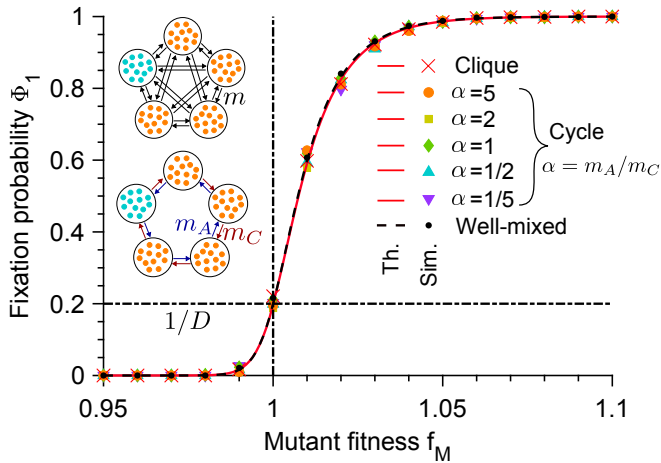}
	\caption{Fixation probability $\Phi_1$ of the mutant type versus mutant fitness $f_M$, for the clique (see upper inset), and for the cycle (see lower inset) with different migration rate asymmetries $\alpha=m_A/m_C$, starting with one fully mutant deme. Data for the well-mixed population is shown as reference, with same total population size and initial number of mutants. Markers are computed over $10^3$ stochastic simulation realizations. Curves represent analytical predictions, Eq.~(\ref{PhiC}) for the cycle and the clique, and Eq.~(\ref{rhoMoran})~\cite{Supplement} for the well-mixed population~\cite{Moran58,Ewens79}. Vertical dash-dotted lines indicate the neutral case $f_M=f_W$, and horizontal dash-dotted lines represent the neutral fixation probability. Parameter values: $D=5$, $K=100$, $f_W=1$, $g_W=g_M=0.1$ in both panels. In simulations, for the clique, $m=10^{-6}$; for the cycle, from top to bottom, $(m_A , m_C) \times 10^6=(1, 5)$; $(1, 2)$; $(1, 1)$; $(2, 1)$; $(5, 1)$.}
	\label{Results_clique}%
\end{figure}

\emph{Clique and cycle.---} In the clique (or island model~\cite{Wright31,Kimura64}), all demes are equivalent and connected to all others with identical migration rates $m$ per individual (Fig. \ref{Results_clique}, upper inset). Starting from one fully mutant deme and $D-1$ fully wild-type demes, the fixation probability $\Phi_1^\textrm{clique}$ of the mutant reads~\cite{Supplement} (proof inspired by \cite{Slatkin81,Traulsen10}):
\begin{equation}
\Phi_1^\textrm{clique}=\frac{1-\gamma}{1-\gamma^D}\mbox{ },
\label{PhiC}
\end{equation}
with 
\begin{equation}
\gamma=\frac{N_W\rho_W}{N_M\rho_M}\,,
\label{gammamt}
\end{equation} 
where $N_W$ (resp. $N_M$) is the deterministic steady-state size of a wild-type (resp. mutant) deme and $\rho_W$ (resp. $\rho_M$) is the fixation probability of a wild-type (resp. mutant) microbe in a mutant (resp. wild-type) deme. This result is independent of migration rate $m$, and Eq.~(\ref{PhiC}) has the exact same form as the fixation probability of a single mutant in a well-mixed population of fixed size $D$ in the Moran model~\cite{Moran58,Ewens79}, but with $\gamma$ playing the role of the ratio $f_W/f_M$, consistently with the formal mapping Eq.~(\ref{mapf}) between our model and that of~\cite{Lieberman05} where $N\rho$ plays the part of fitness. $\Phi_1^\textrm{clique}$ is plotted versus $f_M$ in Fig. \ref{Results_clique}, showing excellent agreement between Eq.~(\ref{PhiC}) and our stochastic simulation results. Moreover, this fixation probability is very close to that in a well-mixed population. We show~\cite{Supplement} that the clique is a slight suppressor of selection, but that modeling migrations as exchanges of individuals and assuming $N_M=N_W$ exactly recovers the well-mixed result, consistently with results on symmetric migrations~\cite{Maruyama70,Maruyama74}. 

Another graph where all demes are equivalent is the cycle. Clockwise and anti-clockwise migrations can have different rates, denoted respectively by $m_C$ and $m_A$ (Fig. \ref{Results_clique}, lower inset). The cycle resembles the circular stepping-stone model~\cite{Maruyama70}, but can feature asymmetric migrations. We show~\cite{Supplement} that the fixation probability $\Phi_1^\textrm{cycle}$ is the same as for the clique, Eq.~(\ref{PhiC}), as corroborated by our simulations, see Fig.~\ref{Results_clique}. Indeed, the cycle is a circulation. In particular, migration rates do not impact $\Phi_1^\textrm{cycle}$. 

\begin{figure*}[hbt]
	\centering
	\includegraphics[width=\textwidth]{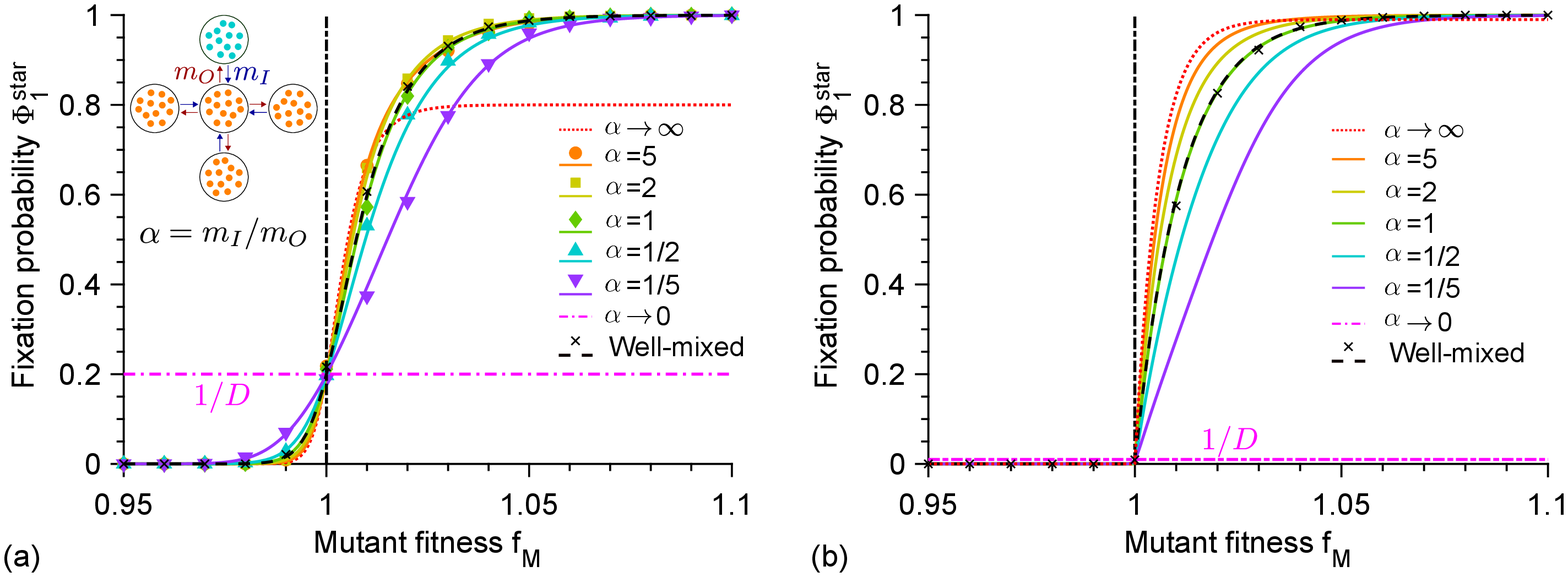}
	\caption{Fixation probability $\Phi_1^\textrm{star}$ of the mutant type in a star graph versus mutant fitness $f_M$, starting with one fully mutant deme chosen uniformly at random, for different migration rate asymmetries $\alpha=m_I/m_O$. Number of demes: $D=5$ (a) and $D=100$ (b). Data for the well-mixed population is shown as reference, with same total population size and initial number of mutants. Markers are computed over $2\times10^3$ stochastic simulation realizations. Curves represent analytical predictions in Eq.~(\ref{PhiStarMT}). Vertical dash-dotted lines indicate the neutral case $f_W=f_M$. Parameter values: $K=100$, $f_W=1$, $g_W=g_M=0.1$. In panel (a), from top to bottom, $(m_I, m_O) \times 10^6=(5, 1)$; $(2, 1)$; $(1, 1)$; $(1, 2)$; $(1, 5)$ in simulations. }
	\label{Results_star}%
\end{figure*} 

\emph{Star.---} In the star, a central node is connected to all others, called leaves. An individual can migrate from a leaf to the center with migration rate $m_I$ and vice-versa with rate $m_O$ (Fig. \ref{Results_star}, inset). The mutant fixation probability can be expressed exactly as a function of $D$, $\alpha=m_I/m_O$ and $\gamma$ defined in Eq.~(\ref{gammamt}) (proof~\cite{Supplement} inspired by~\cite{Broom08}):
\begin{equation}
\Phi_1^\textrm{star}=\frac{(1-\gamma^2)\left[\gamma+\alpha D+\gamma\alpha^2(D-1)\right]}{D\left(\alpha+\gamma\right)\left[1+\alpha\gamma-\gamma^D(\alpha+\gamma)^{2-D}\left(1+\alpha\gamma\right)^{D-1}\right]}\mbox{ }.
\label{PhiStarMT}
\end{equation}

Fig. \ref{Results_star}(a) shows the fixation probability $\Phi_1^\textrm{star}$ of the mutant type for different values of migration asymmetry $\alpha=m_I/m_O$, with very good agreement between Eq.~(\ref{PhiStarMT}) and our simulations. If $\alpha<1$, the star suppresses selection compared to the clique, while for $\alpha>1$ it slightly amplifies selection in some range of mutant fitness $f_M$~\cite{Supplement}. For $\alpha=1$, $\Phi_1^\textrm{star}$ reduces to the fixation probability of the clique, Eq.~(\ref{PhiC})~\cite{Supplement}. Consistently, the star is a circulation for $\alpha=1$. Stronger amplification for $\alpha>1$ is obtained for large $D$ (Fig. \ref{Results_star}(b)). Qualitatively, for large $D$, mutants very likely start in a leaf. If $\alpha$ is large, they often spread to the center, which helps fit mutants take over. Conversely, if $\alpha$ is small, the center often invades the leaves, thus preventing any mutant originating in a leaf from fixing. Results with mutants starting in a specific deme are also shown in~\cite{Supplement}.

Imposing that $\sum_j m_{ij}$ is independent of $i$ amounts to imposing $\alpha=D-1$ in the star~\cite{Supplement}. Then, Eq.~(\ref{PhiStarMT}) reduces to the formula~\cite{Broom08} obtained in the model of~\cite{Lieberman05}, with $\gamma$ in Eq.~(\ref{gammamt}) playing the role of $f_W/f_M$~\cite{Supplement}, as per our general mapping Eq.~(\ref{mapf}). The celebrated amplification property of the star in the large $D$ limit~\cite{Lieberman05,Chalub16} is thus exactly recovered in our model for $\alpha=D-1$.

\begin{figure}[htb]
	\centering
	\includegraphics[width=\columnwidth]{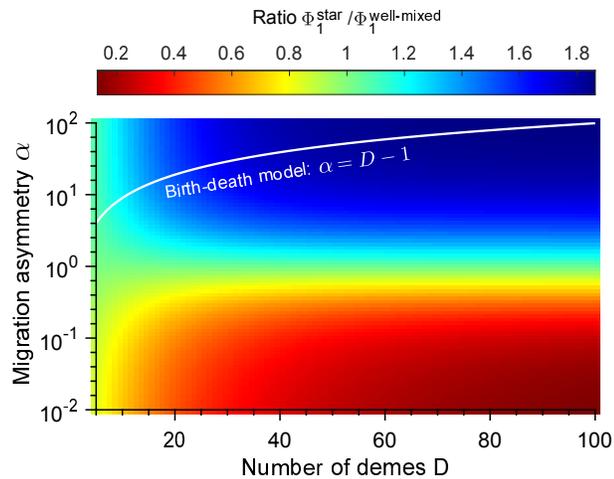}
	\caption{Amplification and suppression properties for the star. Heatmap of the ratio of the fixation probability $\Phi_1^\textrm{star}$ of the mutant type in a star graph to that $\Phi_1^\textrm{well-mixed}$ in a well-mixed population with same total population size and initial number of mutants, versus number $D$ of demes and migration rate asymmetry $\alpha=m_I/m_O$. The star is initialized with one fully mutant deme chosen uniformly at random. Data from analytical formula in Eq.~(\ref{PhiStarMT}) for the star, and in Eq.~(\ref{rhoMoran})~\cite{Supplement} for the well-mixed population. Parameter values: $K=100$, $f_W=1$, $f_M=1.001$, $g_W=g_M=0.1$. }
	\label{Comp_models}%
\end{figure}

While the star is an amplifier for large $D$ in the model of~\cite{Lieberman05}, it can either suppress or an amplify selection, depending on $\alpha$, in our model where $D$ and $\alpha$ are two independent parameters. Fig.~\ref{Comp_models} shows that restricting to $\alpha=D-1$ yields amplification. In models with one individual per node, the star is an amplifier for large $D$ for the Birth-death dynamics (``update rule''), where one individual is chosen to divide and its offspring replaces one of its neighbors~\cite{Lieberman05}, but a suppressor for the death-Birth dynamics (or biased voter model~\cite{Antal06}), where one individual is chosen to die and is replaced by the offspring of one of its neighbors (selection being on division rates in both cases, as denoted by the uppercase ``Birth''~\cite{Hindersin15}, and resulting in global selection in the Birth-death case and local selection in the death-Birth case)~\cite{Frean08,Hadjichrysanthou11,Hindersin15,Allen20}. Consistently, the latter dynamics would yield $\alpha=1/(D-1)$. 

\emph{Comparison to~\cite{Houchmandzadeh11}.---} A model generalizing~\cite{Lieberman05} to graphs where each node contains a deme with fixed population size was introduced in~\cite{Houchmandzadeh11} (see also~\cite{Traulsen05,Campos06,Constable14}). In this model, as in~\cite{Lieberman05}, each elementary event comprises a birth in one deme and a death in another one, yielding Birth-death and death-Birth models that give different results. Rare migrations in our model correspond to strong self-loops (migrations to the original deme) in the model of~\cite{Houchmandzadeh11}. For the star~\cite{Adlam15}, we show~\cite{Supplement} that by matching migration-to-division rate ratios in each deme, both models yield similar simulation results. However, even then, a difference is that death rate (resp. birth rate) is not homogeneous across demes in the Birth-death (resp. death-Birth) models of~\cite{Houchmandzadeh11}, unless migrations are symmetric. Our model allows more realistic choices.

\emph{Discussion.---} We developed a model of spatially structured microbial populations on graphs where migration, birth and death are independent events. We showed that for rare migrations, the star graph continuously transitions between amplifying and suppressing natural selection as migration rate asymmetry is varied. This elucidates the apparent paradox in existing models, where the star, like many random graphs~\cite{Hindersin15}, is an amplifier in the Birth-death dynamics and a suppressor in the death-Birth dynamics~\cite{Frean08,Hadjichrysanthou11,Hindersin15,Allen20}. We found a mapping between our model and that of~\cite{Lieberman05}, under a constraint on migration rates. Models with one individual per node require making specific choices on the microscopic dynamics (``update rule''), which constrain migration rates. By lifting this constraint, our model reconciles and generalizes previous results, showing that migration rate asymmetry is key to whether a given population structure amplifies or suppresses natural selection. This crucial role of migration asymmetry is consistent with the fact that structures with symmetric migrations do not affect fixation probabilities~\cite{Maruyama70,Maruyama74}. 

Birth-death dynamics may be realistic for extreme resource limitation, such that one birth causes one death, while death-Birth dynamics may better model cases where death frees resources, e.g. light for plants~\cite{Alonso06,Frean08}. However, in general, in a microbial population, population size is not strictly fixed, and the order of birth and death events is not set. In our more universal model, results do not hinge on a modeling choice made for microscopic dynamics. Instead, they depend on a quantity that can be directly set or measured in experiments, namely migration rate asymmetry. The differences between Birth-death and death-Birth dynamics are major for mutant fixation probabilities, but also in evolutionary game theory, where spatial structure can promote the evolution of cooperation in the latter case, but not in the former~\cite{Ohtsuki06,Taylor07,Debarre17}. Previous efforts were made to generalize beyond these dynamics by allowing both types of update to occur in given proportions~\cite{Zukewich13,Tkadlec20}. Interestingly, it was recently shown that no general amplification of selection can occur when even a small proportion of death-Birth events occurs~\cite{Tkadlec20}, in contrast to the Birth-death case. Conversely,  in our model, the amplification property of the star graph in the large-size limit is preserved, but for sufficient migration asymmetry. 

While our focus was on mutant fixation probabilities, our model can be employed to investigate fixation times and evolution rate~\cite{Baxter08,Frean13,Hauert14,Hindersin14,Constable14,Lombardo14,Allen15,Hathcock19}. It can also address more complex population structures~\cite{Pavlogiannis18,Allen20}, e.g. motivated by within-host or between-host pathogen dynamics~\cite{Bansept19}. Our study can be extended beyond the regime of rare migrations~\cite{Yagoobi21}, and to models of evolutionary game theory, as well as to diploid organisms~\cite{Nagylaki80,Pulliam88,Barton93,Whitlock97,Whitlock03}. Finally, our work allows direct comparisons with quantitative experiments~\cite{Kassen}. Other experiments could be performed using e.g. microfluidic devices allowing to control flow between different populations~\cite{Oh06}, or microtiter plates where dilutions and migrations can be performed via a liquid-handling robot~\cite{Kryazhimskiy12,Nahum15,France19}. Applications in biotechnology could be envisioned, e.g. amplifying \textit{in vivo} selection in the directed evolution of biomolecules~\cite{Arnold99}.

\textit{Acknowledgments.}--- This project has received funding from the European Research Council (ERC) under the European Union’s Horizon 2020 research and innovation programme (grant agreement No. 851173, to AFB). LM acknowledges funding by a graduate fellowship from École Doctorale Physique en Île-de-France. LM thanks his grandfather, Jean Polard, for inspiration.


\newpage

\onecolumngrid

\beginsupplement


\begin{center}
	\huge{\textbf{Supporting Information}}
\end{center}

\vspace{1cm}

\tableofcontents

\newpage

\section{Fixation probability of neutral mutants} \label{nt}

Consider a graph made of nodes $i$, each associated to a deme with steady-state population size $N_i$, and edges $ij$ where migration rates $m_{ij}$ from deme $i$ to deme $j$ are specified. Further assume that the graph is not disconnected. Consider uniform initial conditions: a mutant has probability $N_i/\sum_j N_j$ to be initially placed in deme $i$.  A neutral mutant then has probability $1/N_i$ to fix in deme $i$ (taking the result for constant population size~\cite{Ewens79}). Let $\Phi_1^{(i)}$ denote the probability that the mutant fixes in the whole metapopulation, starting from a fully mutant deme $i$, all other demes being fully wild-type. The overall fixation probability of one mutant in the metapopulation reads
\begin{equation}
\rho_M=\sum_i \frac{N_i}{\sum_j N_j}\,\frac{1}{N_i}\,\Phi_1^{(i)}=\frac{\sum_i \Phi_1^{(i)}}{\sum_i N_i}\,.
\label{aux0}
\end{equation}
Let us now remark that, because the graph is not disconnected, after a sufficient time, all individuals in the metapopulation are descended from the same deme. If we start from one deme $i$ that is fully mutant and all others that are fully wild-type, this yields
\begin{equation}
\Phi_1^{(i)}+\sum_{j\neq i} \Psi_1^{(j)}=1\,,
\label{aux}
\end{equation}
where $\Psi_1^{(j)}$ is the probability that wild-type individuals from deme $j$ fix in the whole metapopulation. But since the mutant is assumed to be neutral, we have $\Psi_1^{(j)}=\Phi_1^{(j)}$, and thus Eq.~\ref{aux} becomes
\begin{equation}
\sum_{i} \Phi_1^{(i)}=1\,.
\label{aux2}
\end{equation}
Therefore, combining Eqs.~\ref{aux0} and~\ref{aux2}, we obtain
\begin{equation}
\rho_M=\frac{1}{\sum_i N_i}\,,
\label{neutr}
\end{equation}
which is exactly the fixation probability of one neutral mutant in a well-mixed population of size $\sum_i N_i$~\cite{Ewens79}. Thus, provided that the graph is not disconnected, the fixation probability of a neutral mutant under uniform initial conditions is independent of population structure in our model.

In the particular case where all $D$ demes have the same size, i.e. $N_i\equiv N$ does not depend on $i$, then $\rho_M=1/(ND)$, and the average fixation probability starting from one single fully mutant deme under uniform initial conditions is 
\begin{equation}
\Phi_1=\sum_i \frac{N_i}{\sum_j N_j}\,\Phi_1^{(i)}=\frac{\sum_i \Phi_1^{(i)}}{D}=\frac{1}{D}\,.
\label{neutrb}
\end{equation}

\newpage

\section{Fixation probabilities in strongly symmetric graphs}\label{Calculations}

\begin{figure}[htb]
	\includegraphics[scale=0.15]{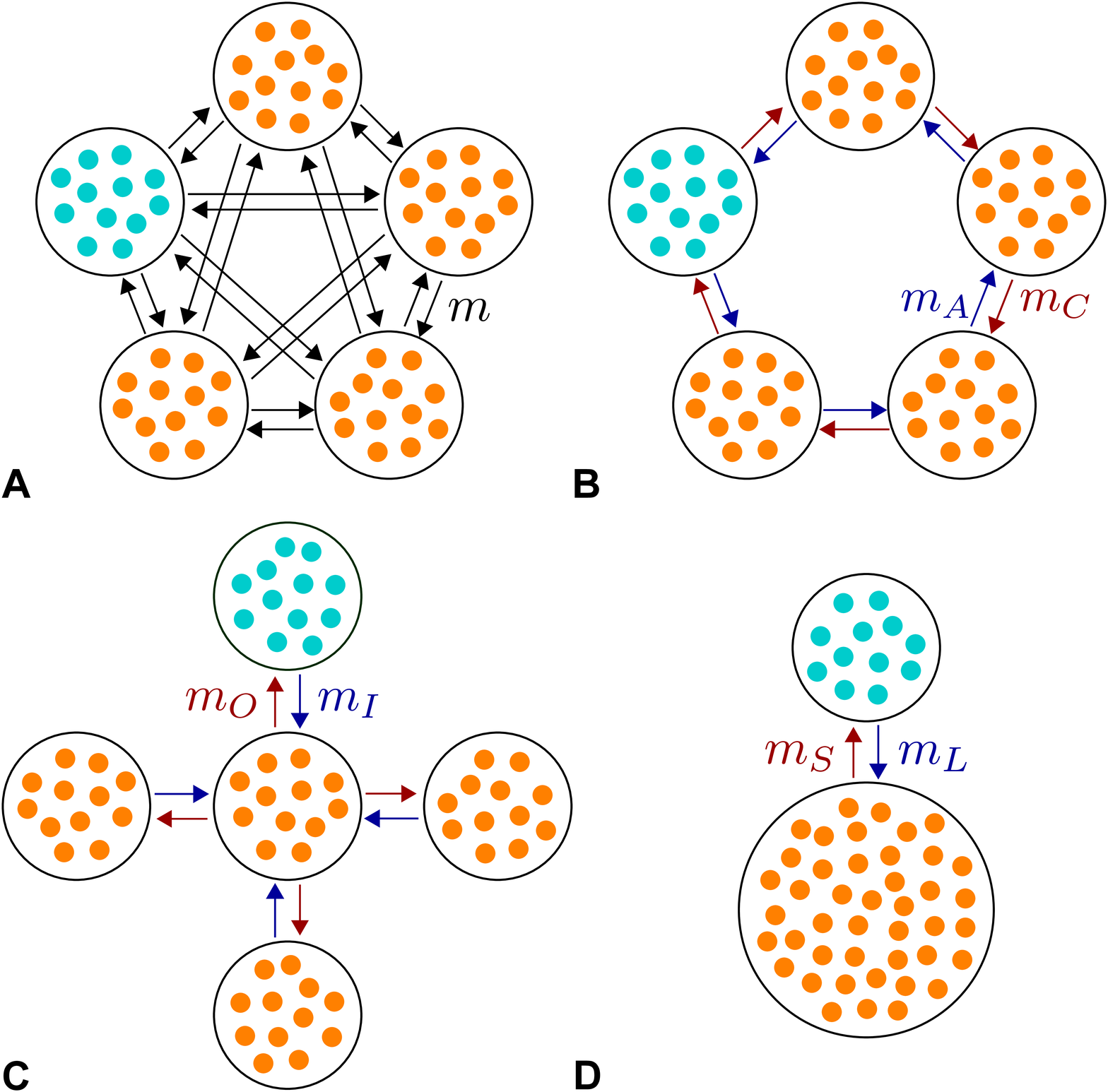}
	\centering
	\caption{{\bf Some population structures.} {\bf A:} Clique. {\bf B:} Cycle. {\bf C:} Star. {\bf D:} Doublet comprising a small deme and a larger deme. Mutants (M) are in blue, wild-type (W) in orange, and the state where  the mutant type has fixed in one deme while all other demes are fully wild-type is represented. Arrows indicate migrations, with the associated rates per individual.}
	\centering
	\label{Struct}
\end{figure}

In this study, we investigate the fate of mutants in the structures shown in Fig \ref{Struct}. The clique, cycle and star are considered in the present section, while the doublet, shown in panel D, is tackled in section \ref{ana_two_pop} as it involves demes with different population sizes.

\subsection{Clique}\label{ana_clique}

\subsubsection{General expression}

Let us consider a population with $D$ demes, structured as a clique, i.e. where migration rates per individual between all demes are equally likely (Fig. \ref{Struct}A). The state of the system can be fully described by the number $i$ of mutant demes. We denote by $m$ the migration rate per individual from one deme to any other deme. Recall that in our model, migration occur between two different demes (no migration can end in the deme where it started). Let us assume that we start from $i$ fully mutant demes and $D-i$ fully wild-type demes. Recall that the wild-type is denoted by $W$ and the mutant by $M$. 

Consider the outcome of a migration event. The number of mutant demes increases by $1$ if an $M$ individual migrates from one of the $i$ mutant demes to one of the $D-i$ wild-type demes, and fixes there. The probability that this occurs upon a migration event thus reads
\begin{equation}
T_i^+=\frac{m N_Mi}{m N_M  i+m N_W (D-i)}\,\frac{D-i}{D-1}\,\rho_M=\frac{N_Mi}{N_M  i+N_W (D-i)}\,\frac{D-i}{D-1}\,\rho_M\,, 
\end{equation}
where 
\begin{equation}
\rho_M=\frac{1-r}{1-r^{N_W}}
\label{rhom}
\end{equation} 
is the fixation probability of a mutant microbe in a wild-type deme in the Moran process~\cite{Moran58,Ewens79} (see section~\ref{EquilibriumSize}) where
\begin{equation}
r=\frac{f_W}{f_M}\,,
\label{defr}
\end{equation} 
and $N_W=K(1-g_W/f_W)$ is the steady-state size of a wild-type deme. Similarly, the number of mutant demes decreases by $1$ if a $W$ individual migrates from one of the $D-i$ wild-type demes to one of the $i$ mutant demes, and fixes there. The probability that this occurs upon a migration event thus reads
\begin{equation}
T_i^-=\frac{N_W(D-i)}{N_M  i+N_W (D-i)}\,\frac{i}{D-1}\,\rho_W\,, 
\end{equation}
where 
\begin{equation}
\rho_W=\frac{1-r^{-1}}{1-r^{-N_M}}
\label{rhow}
\end{equation} 
is the fixation probability of a wild-type microbe in a mutant deme, with $r$ in Eq.~\ref{defr}, and $N_M=K(1-g_M/f_M)$ is the steady-state size of a mutant deme. 

The fixation probability $\Phi_i^\mathrm{clique}$ of the mutant type in a clique of $D$ demes starting with $i$ fully mutant demes satisfies the recurrence relation
\begin{equation}
\left\{
\begin{aligned}
&\Phi_0^\mathrm{clique}=0\\
&\Phi_i^\mathrm{clique}=T_i^+\Phi_{i+1}^\mathrm{clique}+T_i^-\Phi_{i-1}^\mathrm{clique}+(1-T_i^+-T_i^-)\Phi_i^\mathrm{clique}\mbox{ for }1\leq i \leq D-1\\
&\Phi_D^\mathrm{clique}=1\mbox{ },\\
\end{aligned}
\right.
\label{Phi_clique_syst}
\end{equation}
where the second equation follows from distinguishing the different outcomes of the first migration event.
Eq. \ref{Phi_clique_syst} can be solved e.g. as in Ref.~\cite{Traulsen10}, yielding
\begin{equation}
\label{recint}
\Phi_i^\mathrm{clique}=\frac{1+\sum_{k=1}^{i-1}\prod_{j=1}^k\gamma_j}{1+\sum_{k=1}^{D-1}\prod_{j=1}^k\gamma_j}\mbox{ },
\end{equation}
where $\gamma_i=T_i^-/T_i^+=N_W\rho_W/(N_M\rho_M)$. Since here $\gamma_i$ does not depend on the initial number $i$ of mutant demes, the fixation probability $\Phi_i^\mathrm{clique}$ reduces to
\begin{equation}
\Phi_i^\mathrm{clique}=\frac{1-\gamma^i}{1-\gamma^D}\mbox{ },
\label{Phi_SI}
\end{equation}
with 
\begin{equation}
\gamma=\frac{N_W\rho_W}{N_M\rho_M}\,.
\label{gammaDef}
\end{equation}
Note that in the neutral case where $\gamma= 1$, Eq.~\ref{recint} yields $\Phi_i^\mathrm{clique}=i/D$, consistently with Eq.~\ref{neutrb}.

Eq.~\ref{Phi_SI} has the same form as the fixation probability of $i$ mutants in a well-mixed population of $N$ individuals in the Moran model~\cite{Moran58,Ewens79}, namely
\begin{equation}
\rho_{i}=\frac{1-r^{i}}{1-r^{N}}\,,
\label{rhoMoran}
\end{equation}
with $r$ in Eq.~\ref{defr} (note that Eq.~\ref{rhom} corresponds to the case $i=1$). Specifically, Eq.~\ref{rhoMoran} maps to Eq.~\ref{Phi_SI} by replacing $N$ with $D$ and $r$ with $\gamma$. Thus, the clique can be thought of a coarse-grained version of the well-mixed population (see also Ref.~\cite{Slatkin81}): each deme is identically connected to all other demes, just like all individuals are in competition in the well-mixed population. 

To obtain the fixation probability $ \rho_M^\mathrm{clique}$ of a single mutant individual in the clique, one needs to include the fixation of the mutant in one deme before the spread from that deme to the full metapopulation. Thus, it reads
\begin{equation}
\rho_M^\mathrm{clique}=\rho_M \Phi_1^\mathrm{clique} \,,
\label{rhoClique}
\end{equation}
with $\rho_M$ given by Eq.~\ref{rhom} and $\Phi_1^\mathrm{clique}$ by Eq.~\ref{Phi_SI} for $i=1$.

\subsubsection{Expansion for very small mutational effects}
Let $\epsilon$ be such that $f_M=f_W(1+\epsilon)$. Consider the regime where $\epsilon\ll 1$ and $N_W|\epsilon|\ll 1$. Then Eq. \ref{Phi_SI} gives (for $i=1$)
\begin{equation}
\Phi_1^\mathrm{clique}=\frac{1}{D}\left[1+\frac{\epsilon}{2}\left(D-1\right)\left(N_W-1+\frac{2g}{f_W-g}\right)+O(\epsilon^2)\right]\,.
\label{phiClique0}
\end{equation}
Eqs.~\ref{rhoMoran} and \ref{rhoClique} then allow us to show that
\begin{equation}
\frac{\rho_M^\mathrm{clique}}{\rho_{M}^\textrm{well-mixed}}=1-\frac{\epsilon}{2}\left(D-1\right)\left(1-\frac{2g}{f_W-g}\right)+O(\epsilon^2)\,,
\end{equation}
where $\rho_{M}^\textrm{well-mixed}$ is the fixation probability of a mutant in a well-mixed population with $N_W D$ individuals. Therefore, the clique is a suppressor of natural selection in this regime if $g<f/3$. Suppression is all the more important that the degree of subdivision is high, namely the number $D$ of demes (recall that here we compare a clique and a well-mixed population for the same total population size $N_W D$).

\subsubsection{Expansion for relatively small mutational effects}
Next, consider the regime where $\epsilon\ll 1$ and $N_W|\epsilon|\gg 1$ but $N_W\epsilon^2\ll 1$. Then, if $\epsilon>0$,
\begin{equation}
\Phi_1^\mathrm{clique}=1-e^{-N_W\epsilon}\left[1+O(\epsilon)+O(N_W\epsilon^2)\right]\,,
\label{phiClique1Pos}
\end{equation}
which means that fixation is almost certain (recall that if one starts from one single mutant, this holds provided that a mutant has fixed in a deme, which occurs with probability $\rho_{M}=\epsilon+O(\epsilon^2)$, see Eq.~\ref{rhom}). Therefore, Eq.~ \ref{rhoClique} yields $\rho_1^\textrm{clique}=\epsilon+O(\epsilon^2)$, which is equal (to this order) to the fixation probability $\rho_{1,N_WD}^\textrm{well-mixed}$ in a well-mixed population with the same total size $N_W D$. Now, if $\epsilon<0$,
\begin{equation}
\Phi_1^\mathrm{clique}=e^{N_W(D-1)\epsilon}\left[1+O(\epsilon)+O(N_W\epsilon^2)\right]\,.
\label{phiClique1Neg}
\end{equation}
which means that fixation is exponentially suppressed. Thus, Eq.~ \ref{rhoClique} yields $\rho_1^\textrm{clique}=e^{N_WD\epsilon}\left[1+O(\epsilon)+O(N_W\epsilon^2)\right]$, which is equal (to this order) to the fixation probability $\rho_{1}^\textrm{well-mixed}$ in a well-mixed population with the same total size $N_W D$. Hence, in this regime, the fixation probability of a mutant in the clique is very close to that in a well-mixed population with the same total size $N_W D$, and the suppression effect found for extremely small mutational effects is quite restricted.

\subsubsection{Model specialized to symmetric migrations} 
In the clique, migrations are symmetric, i.e. $m_{ij}=m_{ji}=m$ for all $i\neq j$. Let us consider another model, restricted to symmetric migrations, where each migration event is modeled as an exchange between two individuals from two different demes. Let us further neglect the difference between $N_M$ and $N_W$, and assume $N_M=N_W=N$. Upon a given migration event, the probability that the number $i$ of mutant demes increases is 
\begin{equation}
T_i^+=\frac{2i(D-i)}{D(D-1)}\,\rho_M(1-\rho_W)\,, 
\label{symm+}
\end{equation}
and similarly, the probability that $i$ decreases is 
\begin{equation}
T_i^-=\frac{2i(D-i)}{D(D-1)}\,\rho_W(1-\rho_M)\,, 
\label{symm-}
\end{equation}
yielding as above the fixation probability $\Phi_1^\mathrm{clique}$ in Eq.~\ref{Phi_SI} when one starts from one mutant deme, but with 
\begin{equation}
\gamma=\frac{\rho_W(1-\rho_M)}{\rho_M(1-\rho_W)}\,.
\label{gammaSym}
\end{equation}
Eq. \ref{rhoClique} then yields the fixation probability of one single mutant
\begin{equation}
\rho_M^\textrm{clique, sym}=\rho_M \Phi_1^\mathrm{clique}=\frac{1-r}{1-r^{ND}} \,,
\label{rhoCliquesym}
\end{equation}
with $r$ defined in Eq.~\ref{defr}. This is exactly the fixation probability of a mutant in a well-mixed population of fixed size $ND$ in the Moran model (see above).

\subsection{Cycle}\label{ana_ring}

Let us consider a population structured as a cycle with $D$ demes (see Fig. \ref{Struct}B), starting from exactly one fully mutant deme. During the fixation process, this will yield a cluster of $i$ consecutive mutant demes that cannot break. Therefore, in this process, the state of the system can be fully described by the number $i$ of (consecutive) mutant demes. Upon a migration event, the number $i$ of mutant demes increases by $1$ if a $M$ individual from one of the two extremities of the mutant cluster migrates to the neighboring wild-type deme  and fixes there. The probability that this occurs thus reads
\begin{equation}
T_i^+=\frac{(m_C+m_A) N_M}{(m_C+m_A) N_M  i+(m_C+m_A) N_W (D-i)}\,\rho_M\,, 
\label{ring+}
\end{equation}
with $\rho_M$ given by Eq.~\ref{rhom}.
Similarly, the number $i$ of mutant demes decreases by $1$ if a $W$ individual from either of the two wild-type demes surrounding the mutant cluster migrates and fixes in its neighboring mutant deme. The probability that this occurs upon a migration event thus reads
\begin{equation}
T_i^-=\frac{(m_C+m_A) N_W}{(m_C+m_A) N_M  i+(m_C+m_A) N_W (D-i)}\,\rho_W\,.
\label{ring-}
\end{equation}

Thus, the fixation probability $\Phi_i$ of mutation starting with $i$ consecutive mutant demes satisfies Eq. \ref{Phi_clique_syst} with $T_i^+$ and $T_i^-$ given by Eqs.~\ref{ring+} and~\ref{ring-}, which yields the fixation probability in Eq.~\ref{Phi_SI} with $\gamma$ given by Eq.~\ref{gammaDef}. The fixation probability in the cycle, starting from exactly one fully mutant deme, is thus equal to that of the clique with the same number of demes.

\subsection{Star}\label{ana_star}

\subsubsection{General expression}

Let us consider a population structured as a star with $D$ demes (see Fig. \ref{Struct}C). Migrations from each single leaf to the center occur with a rate per individual $m_I$ while migrations from the center to each single leaf occur with a migration rate per individual $m_O$. The state of the system can be fully described by a binary number indicating whether the center is wild-type or mutant and the number $i$ of mutant leaves. 

Upon a given migration event, the probability that the mutant type fixes in the center, if the center is initially wild-type and $i$ leaves are mutant, reads
\begin{equation}
T_{(0,i)\rightarrow(1,i)}=\frac{m_IN_Mi}{m_IN_Mi+m_IN_W(D-1-i)+m_ON_W(D-1)}\rho_M\,,
\end{equation} 
because it happens if migration occurs from a mutant leaf to the center, and the mutant then fixes in the center. Similarly, the probability that the wild-type fixes in the center, if the center is initially mutant and $i$ leaves are mutant, reads
\begin{equation}
T_{(1,i)\rightarrow(0,i)}=\frac{m_IN_W(D-1-i)}{m_IN_Mi+m_IN_W(D-1-i)+m_ON_W(D-1)}\rho_W\,,
\end{equation} 
while the probability that the number of mutant leaves increases by 1 if the center is mutant is
\begin{equation}
T_{(1,i)\rightarrow(1,i+1)}=\frac{m_ON_M(D-1-i)}{m_IN_Mi+m_IN_W(D-1-i)+m_ON_W(D-1)}\rho_M\,,
\end{equation}  
and the probability that the number of mutant leaves decreases by 1 if the center is wild-type is
\begin{equation}
T_{(0,i)\rightarrow(0,i-1)}=\frac{m_ON_W i}{m_IN_Mi+m_IN_W(D-1-i)+m_ON_W(D-1)}\rho_W\,.
\end{equation} 

Let $\Phi_{0,i}^\textrm{star}$ be the fixation probability of the mutant type starting from $i$ fully mutant leaves and a wild-type center. Similarly, let $\Phi_{1,i}^\textrm{star}$ be the fixation probability of the mutant type starting from $i$ fully mutant leaves and a mutant center. The fixation probabilities $\Phi_{0,i}^\textrm{star}$ and $\Phi_{1,i}^\textrm{star}$ satisfy the following recurrence relationship, which is analogous to that in Ref. \cite{Broom08}:
\begin{equation}
\left\{
\begin{aligned}
\Phi_{0,0}^\textrm{star}=\,&\,0 \,,\\
\Phi_{1,i}^\textrm{star}=\,&\,T_{(1,i)\rightarrow(0,i)}\Phi_{0,i}^\textrm{star}+T_{(1,i)\rightarrow(1,i+1)}\Phi_{1,i+1}^\textrm{star}\\&+\left[1-T_{(1,i)\rightarrow(0,i)}-T_{(1,i)\rightarrow(1,i+1)}\right]\Phi_{1,i}^\textrm{star}\mbox{ for }0 \leq i \leq D-2\,,\\
\Phi_{0,i}^\textrm{star}=\,&\,T_{(0,i)\rightarrow(1,i)}\Phi_{1,i}^\textrm{star}+T_{(0,i)\rightarrow(0,i-1)}\Phi_{0,i-1}^\textrm{star}\\&+\left[1-T_{(0,i)\rightarrow(1,i)}-T_{(0,i)\rightarrow(0,i-1)}\right]\Phi_{0,i}^\textrm{star}\mbox{ for }1 \leq i \leq D-1\,,\\
\Phi_{1,D-1}^\textrm{star}=\,&\,1\mbox{ },
\end{aligned}
\right.
\label{Phi_star_1}
\end{equation} 
Employing the expressions of the transition probabilities given above, the system \ref{Phi_star_1} can be rewritten as:
\begin{equation}
\left\{
\begin{aligned}
&\Phi_{0,0}^\textrm{star}=0\,, \\
&\Phi_{1,i}^\textrm{star}=\Phi_{1,i-1}^\textrm{star}+\Gamma_1(\Phi_{1,i-1}^\textrm{star}-\Phi_{0,i-1}^\textrm{star})\mbox{ for }1 \leq i \leq D-1\,,\\
&\Phi_{0,i}^\textrm{star}=\frac{1}{1+\Gamma_0}\Phi_{1,i}^\textrm{star}+\frac{\Gamma_0}{1+\Gamma_0}\Phi_{0,i-1}^\textrm{star}\mbox{ for }1 \leq i \leq D-1\,,\\
&\Phi_{1,D-1}^\textrm{star}=1\mbox{ },
\end{aligned}
\right.
\label{Phi_star_2}
\end{equation}
where $\Gamma_1=\gamma\,m_I/m_O$ and $\Gamma_0=\gamma\,m_O/m_I$, with $\gamma$ given in Eq.~\ref{gammaDef}. Solving the system \ref{Phi_star_2} yields
\begin{equation}
\left\{
\begin{aligned}
&\Phi_{0,0}^\textrm{star}=0\,, \\
&\Phi_{1,i}^\textrm{star}=\frac{-1+\Gamma_1\left[-1+(1+\Gamma_0)\left(\frac{\Gamma_0(1+\Gamma_1)}{1+\Gamma_0}\right)^i\right]}{-1+\Gamma_1\left[-1+(1+\Gamma_0)\left(\frac{\Gamma_0(1+\Gamma_1)}{1+\Gamma_0}\right)^{D-1}\right]}\mbox{ for }0 \leq i \leq D-2\,,\\
&\Phi_{0,i}^\textrm{star}=\frac{(1+\Gamma_1)\left[-1+\left(\frac{\Gamma_0(1+\Gamma_1)}{1+\Gamma_0}\right)^i\right]}{-1+\Gamma_1\left[-1+(1+\Gamma_0)\left(\frac{\Gamma_0(1+\Gamma_1)}{1+\Gamma_0}\right)^{D-1}\right]}\mbox{ for }1 \leq i \leq D-1\,,\\
&\Phi_{1,D-1}^\textrm{star}=1\mbox{ }.
\end{aligned}
\right.
\label{Phi_star}
\end{equation}
In particular, the fixation probability of the mutant type starting from one fully mutant center and all leaves fully wild-type reads
\begin{equation}
\Phi_{1,0}^\textrm{star}=\frac{1-\gamma^2}{1+\alpha\gamma-\gamma(\alpha+\gamma)\left(\frac{\gamma(1+\alpha\gamma)}{\alpha+\gamma}\right)^{D-1}}\mbox{ },
\label{Phi10}
\end{equation} 
where 
\begin{equation}
\alpha=\frac{m_I}{m_O}\,.
\end{equation}  
The fixation probability of the mutant type starting from one fully mutant leaf and all other demes fully wild-type reads
\begin{equation}
\Phi_{0,1}^\textrm{star}=\frac{\frac{\alpha}{\alpha+\gamma}(1+\alpha\gamma)\left(1-\gamma^2\right)}{1+\alpha\gamma-\gamma(\alpha+\gamma)\left(\frac{\gamma(1+\alpha\gamma)}{\alpha+\gamma}\right)^{D-1}}=\frac{\alpha}{\alpha+\gamma}(1+\alpha\gamma)\Phi_{1,0}^\textrm{star}\mbox{ }.
\label{Phi01}
\end{equation}
The probability that the mutant type fixes, starting from a mutant deme that can be any deme of the star with equal probability, can then be expressed as 
\begin{align}
\Phi_1^\textrm{star}&=\frac{1}{D}\Phi_{1,0}^\textrm{star}+\frac{D-1}{D}\Phi_{0,1}^\textrm{star}=\frac{\gamma+\alpha D+\gamma\alpha^2(D-1)}{D\left(\alpha+\gamma\right)}\Phi_{1,0}^\textrm{star}\nonumber\\
&=\frac{\gamma+\alpha D+\gamma\alpha^2(D-1)}{D\left(\alpha+\gamma\right)}\frac{1-\gamma^2}{1+\alpha\gamma-\gamma(\alpha+\gamma)\left(\frac{\gamma(1+\alpha\gamma)}{\alpha+\gamma}\right)^{D-1}}\mbox{ }.
\label{PhiStar}
\end{align}
which can be rewritten as Eq.~\ref{PhiStarMT} in the main text.

Importantly, Eqs. \ref{Phi10}, \ref{Phi01} and \ref{PhiStar} are exactly equivalent to the formula given in Ref.~\cite{Broom08} if the following substitutions are made: $\gamma \rightarrow f_W/f_M$ and $\alpha \rightarrow D-1$ (bearing in mind that in the notations of Ref.~\cite{Broom08}, $D-1$ is called $n$ and $f_W/f_M$ is called $1/r$).

\subsubsection{Expansion for very small mutational effects}
Let $\epsilon$ be such that $f_M=f_W(1+\epsilon)$. For uniform initialization, consider the regime where $\epsilon\ll 1$ and $N_W|\epsilon|\ll 1$. Then Eq. \ref{PhiStar} yields
\begin{equation}
\Phi_1^\textrm{star}=\frac{1}{D}\left[1+\epsilon\,\frac{\alpha(D-1)\left[\alpha(D-2)+2\right]}{(\alpha+1)\left[\alpha(D-1)+1\right]}\left(N_W-1+\frac{2g}{f-g}\right)+O(\epsilon^2)\right]\,.
\label{phiStar0}
\end{equation}
Comparing Eqs.~\ref{phiClique0} and \ref{phiStar0} yields
\begin{equation}
\frac{\Phi_1^\mathrm{star}}{\Phi_1^\mathrm{clique}}=1+\frac{\epsilon}{2}\left(N_W-1+\frac{2g}{f-g}\right)(D-1)\frac{(\alpha-1)\left[\alpha(D-3)+1\right]}{(\alpha+1)\left[\alpha(D-1)+1\right]}+O(\epsilon^2)\,.
\label{ratioStar0}
\end{equation}
Assuming $D>2$, the first-order term in Eq.~\ref{ratioStar0} has the same sign as $\epsilon(\alpha-1)$. Thus, in this regime, the star is an amplifier of selection with respect to the clique for $\alpha>1$, and a suppressor for $\alpha<1$. Furthermore, for $D>3$ (and integer), the function
\begin{equation}
F:\alpha\mapsto\frac{(\alpha-1)\left[\alpha(D-3)+1\right]}{(\alpha+1)\left[\alpha(D-1)+1\right]}\,,
\label{theF}
\end{equation}
increases with $\alpha$ for $\alpha>0$, which entails that, for very small mutational effects, the strongest amplification is obtained for $\alpha\gg 1$, where $F(\alpha)\to 1$. Conversely, if $\alpha\ll 1$ and $\alpha\ll 1/D$, Eq.~\ref{phiStar0} yields
\begin{equation}
\Phi_1^\textrm{star}=\frac{1}{D} +\epsilon\,\frac{2\alpha(D-1)}{D}\left(N_W-1+\frac{2g}{f-g}\right)+O(\epsilon^2)\,.
\label{phiStar0smallm}
\end{equation}
In particular, the coefficient of the first-order term in $\epsilon$ becomes very small if $\alpha\ll 1/(2N_W)$, meaning that for such small values of $\alpha$, we expect a strong suppression of selection, with a fixation probability that becomes independent of $\epsilon$ and flat (for very small mutational effects  $\epsilon$).

\subsubsection{Expansion for relatively small mutational effects}
Next, consider the regime where $\epsilon\ll 1$ and $N_W|\epsilon|\gg 1$ but $N_W\epsilon^2\ll 1$. Then, if $\epsilon>0$, Eq. \ref{PhiStar} yields
\begin{equation}
\Phi_1^\textrm{star}=1-e^{-N_W\epsilon}\frac{D+\alpha^2-1}{\alpha D}\left[1+O(\epsilon)+O(N_W\epsilon^2)\right]\,,
\label{phiStar1Pos}
\end{equation}
which gives, employing Eq.~\ref{phiClique1Pos},
\begin{equation}
\frac{\Phi_1^\mathrm{star}}{\Phi_1^\mathrm{clique}}=1+e^{-N_W\epsilon}(\alpha-1)\frac{D-\alpha-1}{\alpha D}\left[1+O(\epsilon)+O(N_W\epsilon^2)\right]\,.
\label{ratioStar1Pos}
\end{equation}
Thus, in this case, assuming $D>2$, we have $\Phi_1^\mathrm{star}<\Phi_1^\mathrm{clique}$ if $\alpha<1$ or $\alpha>D-1$, whereas  $\Phi_1^\mathrm{star}>\Phi_1^\mathrm{clique}$ if $1<\alpha<D-1$. Now if $\epsilon<0$, Eq. \ref{PhiStar} yields
\begin{equation}
\Phi_1^\textrm{star}=\frac{1+\alpha^2(D-1)}{D\,\alpha^{D-1}}e^{N_W(D-1)\epsilon}\left[1+O(\epsilon)+O(N_W\epsilon^2)\right]\,,
\label{phiStar1Neg}
\end{equation}
which gives, employing Eq.~\ref{phiClique1Neg},
\begin{equation}
\frac{\Phi_1^\mathrm{star}}{\Phi_1^\mathrm{clique}}=\frac{1+\alpha^2(D-1)}{D\,\alpha^{D-1}}\left[1+O(\epsilon)+O(N_W\epsilon^2)\right]\,.
\label{ratioStar1Neg}
\end{equation}
Then, assuming $D>3$, we have $\Phi_1^\mathrm{star}>\Phi_1^\mathrm{clique}$ if $\alpha<1$ while $\Phi_1^\mathrm{star}<\Phi_1^\mathrm{clique}$ if $\alpha>1$. 

Combining results for $\epsilon>0$ and $\epsilon<0$ in this regime, as well as results obtained for very small mutational effects above, we find that the star is a suppressor of selection compared to the clique for $\alpha<1$, an amplifier of selection for $1<\alpha<D-1$, and a transient amplifier of selection for $\alpha>D-1$. Indeed, in the latter case, one switches from amplification to suppression as $\epsilon$ is increased. Specifically, there is amplification for $\epsilon<0$ satisfying $\epsilon\ll 1$ and $N_W|\epsilon|\gg 1$ but $N_W\epsilon^2\ll 1$, and whatever the sign of $\epsilon$ in the regime where $\epsilon\ll 1$ and $N_W|\epsilon|\ll 1$, but there is suppression for $\epsilon>0$ satisfying $\epsilon\ll 1$ and $N_W|\epsilon|\gg 1$ but $N_W\epsilon^2\ll 1$.

Interestingly, in the regime of very small mutational effects where $\epsilon\ll 1$ and $N_W|\epsilon|\ll 1$, we showed that the strongest amplification is obtained in the limit $\alpha\gg 1$, but now we find that in this case, amplification is only transient. Our expansions show that universal amplification can exist only if $1<\alpha\leq D-1$. In this case, for $\epsilon\ll 1$ and $N_W|\epsilon|\ll 1$, the strongest amplification is expected for $\alpha=D-1$ (because $F$ increases with $\alpha$, see Eq.~\ref{theF}), and we then have $F(D-1)=(D-2)^3/\left\{D\left[D\left(D-2\right)+2\right]\right\}$ so that Eq.~\ref{ratioStar0} then yields
\begin{equation}
\frac{\Phi_1^\mathrm{star}}{\Phi_1^\mathrm{clique}}=1+\frac{\epsilon}{2}\left(N_W-1+\frac{2g}{f-g}\right)\frac{(D-1)(D-2)^3}{D\left[D\left(D-2\right)+2\right]}+O(\epsilon^2)\,.
\label{ratioStar0mSpe}
\end{equation}
If $N_W\gg 1$ and $D\gg 1$, this gives a prefactor of the first order term in $\epsilon$ of order $N_W D/2$, which can yield a large amplification, but recall that this is restricted to $N_W|\epsilon|\ll 1$.

\subsubsection{Expansion for extremely asymmetric migrations} 
So far we have considered expansions in selection strengths, and then in some regimes, analyzed extremely asymmetric migrations. However, the order of limits matters and our previous discussions are limited to specific regimes in terms of selection strength. If $\alpha\to 0$, Eq. \ref{PhiStar} yields
\begin{equation}
\Phi_1^\textrm{star}=\frac{1}{D}+O(\alpha)\,,
\end{equation}
which demonstrates that for small $m$ the star is a very strong suppressor of selection, with all mutations becoming effectively neutral (once they have fixed in a deme). Note that this is consistent with our result for very small mutational effects, see Eq.~\ref{phiStar0smallm} and the discussion just below. 

If $\alpha\to \infty$, and in particular assuming $\alpha\gg D$, Eq. \ref{PhiStar} yields
\begin{equation}
\Phi_1^\textrm{star}=\frac{D-1}{D}\frac{1-\gamma^2}{1-\gamma^{2(D-1)}}+O(\alpha^{-1})\,,
\end{equation}
and  for $\gamma=N_W\rho_W/(N_M\rho_M)\to 0$, which occurs when $f_M\gg f_W$, we have $\Phi_1^\textrm{star} \to (D-1)/D$, which confirms that amplification can only be transient in this case, since for the clique, we have $\Phi_1^\textrm{star} \to 1$ in this limit. The simple expression $\Phi_1^\textrm{star} \to (D-1)/D$ is due to the fact that in the $\alpha\to\infty$, mutants in the center cannot fix even if they are very fit, while those in the leaves fix easily. If in addition $D\gg 1$ (and thus  $\alpha\gg D\gg 1$), then
\begin{equation}
\Phi_1^\textrm{star}\approx\frac{1-\gamma^2}{1-\gamma^{2(D-1)}}+O(\alpha^{-1})\,,
\label{asympt1}
\end{equation}
which is formally reminiscent of the fixation probability for the star in the model of Ref.~\cite{Lieberman05} with Birth-death dynamics in the limit $D\to\infty$ (see also Ref.~\cite{Broom08}), where $\gamma$ replaces the ratio of fitnesses (as is the case in Eq.~\ref{PhiC} for the clique, which is formally reminiscent of the fixation probability for the well-mixed population). Thus, if $\alpha\to\infty$ such that $\alpha\gg D\gg 1$, the star can become a universal amplifier of selection with respect to the clique, and has a fixation probability identical to that of the clique but with $\gamma$ replaced by $\gamma^2$, which demonstrates amplification, in the same way as in the model of Ref.~\cite{Lieberman05}. However, this is restricted to the particular regime $\alpha\to\infty$ such that $\alpha\gg D\gg 1$.

In the Birth-death model~\cite{Lieberman05}, the star satisfies $m_O=1/(D-1)$ and $m_I=1$, so that $\alpha=D-1$ (see section \ref{map} for a more general mapping between our model and that of Ref.~\cite{Lieberman05}). Let us thus consider the specific case where $\alpha=D-1$. If $\alpha\to \infty$ (which implies $D\to\infty$), Eq. \ref{PhiStar} yields
\begin{equation}
\Phi_1^\textrm{star}=\frac{1-\gamma^2}{1-\gamma^{2(D-1)}e^{(1-\gamma^2)/\gamma}}+O(\alpha^{-1})\,,
\end{equation}
which has the exact same form as the rigorous asymptotic expression~\cite{Chalub16} for $D\to\infty$ of the fixation probability in a star in the Birth-death model of Ref.~\cite{Lieberman05}. This is consistent with the fact that Eqs. \ref{Phi10}, \ref{Phi01} and \ref{PhiStar} are exactly equivalent to the formula given in Ref.~\cite{Broom08} with $\gamma \rightarrow f_W/f_M$ and $\alpha \rightarrow D-1$ (see above). Note that the rigorous asymptotic expression from Ref.~\cite{Chalub16} is slightly different from the better known expression that has the same form as Eq.~\ref{asympt1}, which holds for $\alpha\gg D\gg 1$ in our model.

\subsubsection{Additional results for the star} 
\begin{figure}[htb]
	\centering
	\includegraphics[width=\textwidth]{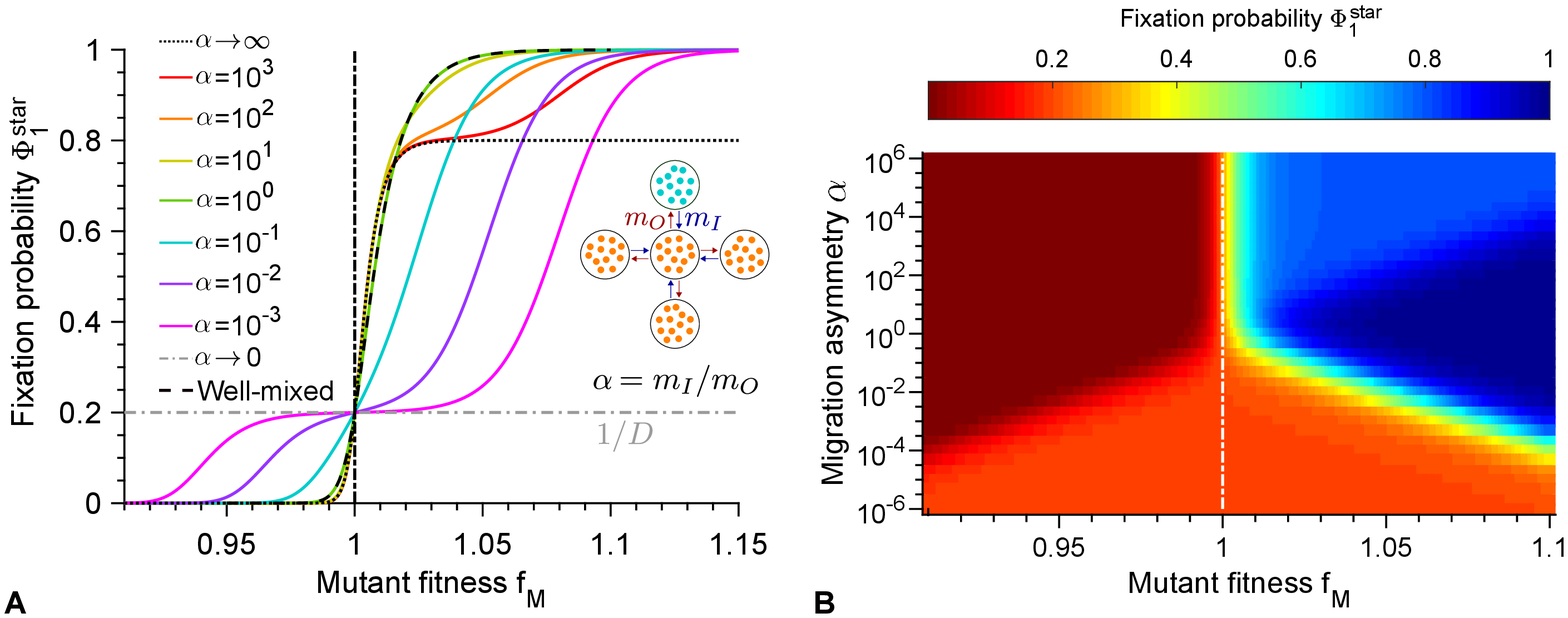}
	\caption{{\bf Fixation probability for the star.} {\bf A:} Fixation probability $\Phi_1^\textrm{star}$ of mutants in a star graph versus mutant fitness $f_M$, starting with one fully mutant deme chosen uniformly at random, with different migration rate asymmetries $\alpha=m_I/m_O$, complementing those shown in Fig.~\ref{Results_star}A. Data for the well-mixed population is shown as reference, with same total population size and initial number of mutants. Curves represent analytical predictions in Eq.~\ref{PhiStar}. {\bf B:} Heatmap of the same fixation probability $\Phi_1^\textrm{star}$ shown versus mutant fitness $f_M$ and migration rate asymmetry $\alpha=m_I/m_O$. Parameter values in both panels: $D=5$, $K=100$, $f_W=1$, $g_W=g_M=0.1$. Vertical dash-dotted lines represent the neutral case $f_W=f_M$. }
	\label{Results_star_more}%
\end{figure}

In Fig.~\ref{Results_star_more}, we show results for the fixation probability in the star graph that complement those shown in Fig.~\ref{Results_star}.

\begin{figure}[htb]
	\centering
	\includegraphics[width=\textwidth]{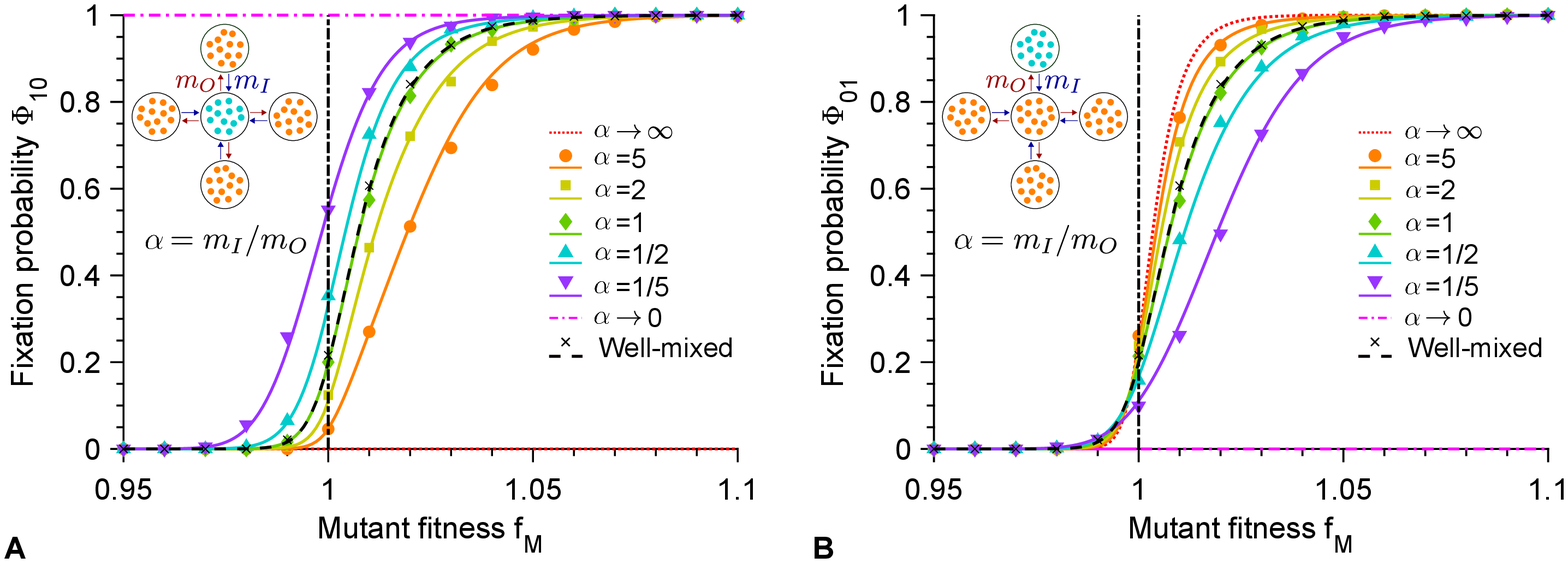}
	\caption{{\bf Fixation probability for the star, starting from a mutant center or leaf.} {\bf A:} Fixation probability $\Phi_{10}$ of mutants in a star graph versus mutant fitness $f_M$, starting with a fully mutant center, with different migration rate asymmetries $\alpha=m_I/m_O$. Markers are computed over $10^3$ stochastic simulation realizations. Curves represent analytical predictions in Eq.~\ref{Phi10}. {\bf B:} Fixation probability $\Phi_{01}$ of mutants in a star graph versus mutant fitness $f_M$, starting with a fully mutant leaf, with different migration rate asymmetries $\alpha=m_I/m_O$. Markers are computed over $10^3$ stochastic simulation realizations. Curves represent analytical predictions in Eq.~\ref{Phi01}. In both panels, vertical dash-dotted lines represent the neutral case $f_W=f_M$. Data for the well-mixed population is shown as reference, with same total population size and initial number of mutants. Parameter values in both panels: $D=5$, $K=100$, $f_W=1$, $g_W=g_M=0.1$. From top to bottom, $(m_I, m_O) \times 10^6=(5, 1)$; $(2, 1)$; $(1, 1)$; $(1, 2)$; $(1, 5)$ in simulations. }
	\label{Star_Center_Leaf}%
\end{figure}

\begin{figure}[htb]
	\centering
	\includegraphics[width=\textwidth]{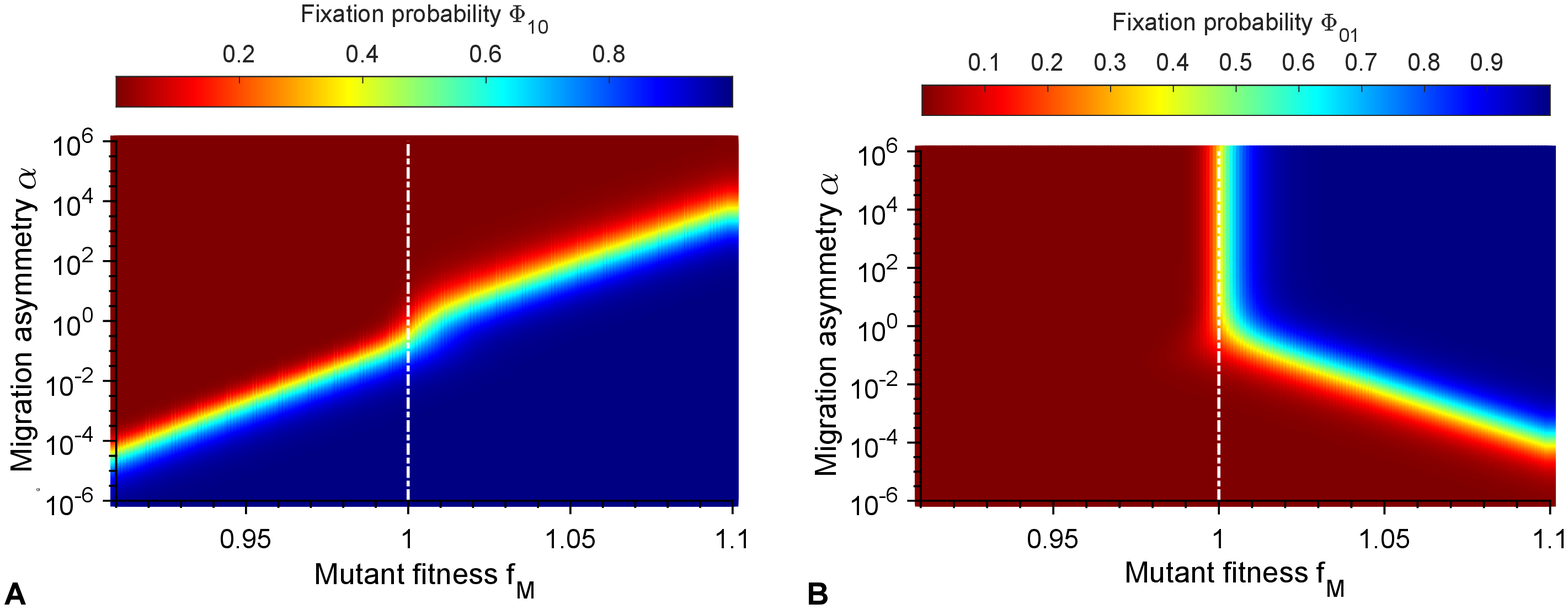}
	\caption{{\bf Heatmaps of the fixation probability for the star, starting from a mutant center or leaf.} {\bf A:} Heatmap of the fixation probability $\Phi_{10}$ of mutants in a star graph starting with a fully mutant center, shown versus mutant fitness $f_M$ and migration rate asymmetry $\alpha=m_I/m_O$. {\bf B:} Similar heatmap but for the fixation probability $\Phi_{01}$ of mutants in a star graph starting with a fully mutant leaf. In both panels, vertical dash-dotted lines represent the neutral case $f_W=f_M$. Parameter values in both panels: $D=5$, $K=100$, $f_W=1$, $g_W=g_M=0.1$.}
	\label{Star_Center_Leaf_HM}%
\end{figure}

In this work, we usually start from a mutant deme chosen uniformly at random, which is realistic for spontaneous mutations. However, the initial position of the mutant~\cite{Lieberman05}, and the degree of the node where it starts~\cite{Antal06}, can strongly impact its fate. Thus, in Figs.~\ref{Star_Center_Leaf} and~\ref{Star_Center_Leaf_HM}, we show results when the initial mutant deme is either the center or a leaf. These results illustrate the strong impact of mutant initial position. 

\clearpage

\section{Comparison with the model of Ref.~\cite{Lieberman05}}
\label{map}

In Ref.~\cite{Lieberman05}, a model where each of the $N$ nodes of a graph is occupied by a single individual was introduced. Replacement probabilities $w_{ij}$ from node $i$ to node $j$ are defined along each edge $ij$ of the graph. At each elementary step, an individual (say the one on node $i$) is selected for division, with probability proportional to fitness $f_i$, and its offspring replaces the individual on node $j$ with probability $w_{ij}$. This dynamics, which became known as the Birth-death dynamics~\cite{Kaveh15, Hindersin15, Pattni15} or biased invasion process~\cite{Antal06,Houchmandzadeh11}, thus allows to always maintain exactly one individual on each node. An important constraint stemming from the definition of the model is 
\begin{equation}
\sum_{j=1}^N w_{ij}=1\,, 
\label{rightstoch}
\end{equation}
because the offspring of individual $i$ has to end up somewhere. In other words, the matrix of replacement probabilities is right-stochastic. Note that self-loops where the offspring stays on the same node (corresponding to $w_{ii}>0$) were not considered in the initial description of the model but can be added (see e.g.~\cite{Adlam15}). The probability $P_{i\rightarrow j}$ that, at a given elementary step, the offspring from node $i$ replaces the individual in node $j$ is given by
\begin{equation} \label{eq:BD Pij}
P_{i\rightarrow j} = \frac{f_i}{\sum_{k=1}^N f_k} w_{ij}\,,
\end{equation}
i.e. the probability that the individual on node $i$ is selected for division, multiplied by the probability that its offspring replaces the individual in node $j$. Note that, since exactly one replacement occurs per elementary step, $\sum_{i,j}P_{i\rightarrow j}=1$, and that Eq.~\ref{eq:BD Pij} satisfies this normalization constraint because Eq.~\ref{rightstoch} holds. Using Eq.~\ref{rightstoch}, we can rewrite Eq.~\ref{eq:BD Pij} as
\begin{equation} \label{eq:BD Pij 2}
P_{i\rightarrow j} = \frac{f_i  w_{ij}}{\sum_{k,\,l} f_k w_{kl}}\,,
\end{equation}
which will be convenient for our comparison.

In our coarse-grained model, upon each migration event, the individual that migrated from deme $i$ to deme $j$ (with migration rate $m_{ij}$ per individual) may fix with probability $\rho_i$. In particular, the probability $P_{i}^\mathrm{mut}$ that a specific deme $i$ becomes mutant upon one given migration event while it was wild-type before reads
\begin{equation} 
P_{i}^\mathrm{mut} = \frac{\sum_{k M} N_k m_{ki}}{\sum_{k,\,j} N_km_{kj}} \rho_{i}= \frac{ N_M\sum_{k M} m_{ki}}{N_M\sum_{k M,\,j} m_{kj} +N_W\sum_{k W,\,j} m_{kj}}\rho_{M} \,,
\end{equation}
where $\sum_{k M}$ denotes a sum over mutant ($M$) demes indexed by $k$. In the last term we discriminated over mutant and wild-type demes and employed the fact that all demes have the same carrying capacity $K$, resulting in steady-state sizes $N_M$ for mutant demes and $N_W$ for wild-type demes, and denoted by $\rho_M$ the fixation probability of a mutant in a wild-type deme, following our usual convention. Here, we have considered a probability upon a migration event, but migration events change the makeup of the population only if fixation ensues. To compare to the model of Ref.~\cite{Lieberman05}, let us instead focus only on the migration events that result into fixation. The probability $\mathcal{P}_{i\rightarrow j}$ that, upon such a successful migration event, an individual coming from deme $i$ fixes in deme $j$ reads
\begin{equation} \label{Pij us}
\mathcal{P}_{i\rightarrow j} = \frac{N_im_{ij}\rho_{i}}{\sum_{k,\,l}N_km_{kl}\rho_{k}}\,.
\end{equation}
Note that it satisfies $\sum_{i,\,j}\mathcal{P}_{i\rightarrow j}=1$, as a fixation occurs at each successful migration event.

Eqs.~\ref{eq:BD Pij 2} and~\ref{Pij us} have the same form, with $N_i\rho_{i}$ in our model playing the part of $f_i$ in the model of Ref.~\cite{Lieberman05}. An important difference is that in our model, the $m_{ij}$ (which are migration rates, not migration probabilities) do not need to satisfy the constraint in Eq.~\ref{rightstoch} and are independent. Our model is thus less constrained than that of Ref.~\cite{Lieberman05}. Note that an alternative dynamics removing this constraint was discussed in Ref.~\cite{Lieberman05}, but then very rarely considered in the literature~\cite{Pattni15}. Note also that the fixation probability $\rho_i$ involves the fitness of $i$ and that of the type that is replaced (say $j$), but because we always work with just two types, this dependence can be ignored without losing generality.

For the clique and for the cycle, we have found that the fixation probability, given by Eq.~\ref{Phi_SI}, has the same form as that for the well-mixed population, but with $\gamma=N_W\rho_W/(N_M\rho_M)$ playing the part of $f_W/f_M$. This is perfectly consistent with the mapping described here, with $N_i\rho_{i}$ in our model playing the part of $f_i$. Note that the constraint on migration rates does not come into play here since Eq.~\ref{Phi_SI} is independent of migration rates.

For the star, we already noted that the fixation probabilities given in Eqs. \ref{Phi10}, \ref{Phi01} and \ref{PhiStar} are exactly equivalent to the formula given in Ref.~\cite{Broom08} if the following substitutions are made: $\gamma=N_W\rho_W/(N_M\rho_M) \rightarrow f_W/f_M$ and $\alpha=m_I/m_O \rightarrow D-1$ (bearing in mind that in the notations of Ref.~\cite{Broom08}, $D-1$ is called $n$ and $f_W/f_M$ is called $1/r$). Again, this is perfectly consistent with the mapping described here, with $N_i\rho_{i}$ in our model playing the part of $f_i$. In addition, we have to impose a specific value of $\alpha$, namely $\alpha=D-1$, in order to get back the result of Ref.~\cite{Broom08}. This is because, for a star graph with no self-loops, only two different migration rates can exist, $m_O$ from center to leaf and $m_I$ from leaf to center, given the symmetries of this graph. Imposing that $\sum_j m_{ij}$ is independent on $i$, i.e. that all nodes have the same total emigration rate, which is a weaker form of the constraint in Eq.~\ref{rightstoch} (because no normalization is required on migration rates), then yields $\alpha=D-1$. So the extra constraint in the mapping between the two models for the star stems from the requirement that Eq.~\ref{rightstoch} be satisfied in the model of Ref.~\cite{Lieberman05}. This also means that for $\alpha=D-1$, our results are formally the same as in Ref.~\cite{Broom08} and in the model of Ref.~\cite{Lieberman05} and that we then find the exact same amplification properties for the star. But this exact correspondence is restricted to a very particular value of $\alpha$.

\section{Generalized circulation theorem} \label{circt}

Here, we extend the circulation theorem from Ref.~\cite{Lieberman05} to our model. Consider a metapopulation on a graph $G$ with a set of nodes $\bm{V}$ where all $D$ demes have the same carrying capacity. A graph with migration rates per individual $m_{ij}$ from node $i$ to node $j$ is a circulation if and only if for all $i$, 
\begin{equation}
\sum_{j\in\bm{V}} m_{ij}=\sum_{j\in\bm{V}} m_{ji}\,,
\label{def_circ}
\end{equation}
which means that the total rate of migrations leaving $i$ is equal to the total rate of migrations arriving in $i$. We will show that the fixation probability starting from $P$ fully mutant demes is the same as for the clique, i.e. is given by Eq.~\ref{Phi_SI}, if and only if the graph $G$ is a circulation. Let us denote by $\bm{P}$ the ensemble of fully mutant demes and by $P$ its cardinal.

Following  the proof of the circulation theorem given in Ref.~\cite{Lieberman05}, we will demonstrate that the following are equivalent:\\
(1) G is a circulation.\\
(2) $P$ performs a random walk with forward bias $\gamma^{-1}$, with $\gamma=N_W\rho_W/(N_M\rho_M)$, and absorbing states $\{0,D\}$.\\
(3) The fixation probability starting from $P$ fully mutant demes is the same as for the clique, i.e. is given by Eq.~\ref{Phi_SI}.\\
(4) The probability that, starting from any $P$ fully mutant demes, a mutant such that $N_W\rho_W/(N_M\rho_M)=\gamma$ eventually fixes in $P'$ mutant demes is given by
\begin{equation}
\Phi(\gamma, G, P,P')=\frac{1-\gamma^P}{1-\gamma^{P'}}\,.
\label{(4)}
\end{equation}

First we show that $(1) \Rightarrow (2)$, in a similar way as in Ref.~\cite{Lieberman05}. For this, let $\delta_+(\bm{P})$ (resp. $\delta_-(\bm{P})$) be the probability that the number of mutant demes increases by one (resp. decreases by one). We have
\begin{equation}
\frac{\delta_-(\bm{P})}{\delta_+(\bm{P})}=\frac{\sum_{i\in\bm{V}\setminus\bm{P},\,\, j\in\bm{P}} N_W m_{ij} \rho_W}{\sum_{i\in\bm{P},\,\, j\in\bm{V}\setminus\bm{P}} N_M m_{ij} \rho_M}=\gamma\,\, \frac{\sum_{i\in\bm{V}\setminus\bm{P},\,\, j\in\bm{P}} m_{ij}}{\sum_{i\in\bm{P},\,\, j\in\bm{V}\setminus\bm{P}}  m_{ij}}=\gamma\,\, \frac{\sum_{ i\in\bm{P},\,\,j\in\bm{V}\setminus\bm{P}} m_{ji}}{\sum_{i\in\bm{P},\,\, j\in\bm{V}\setminus\bm{P}}  m_{ij}}
\label{ratDelta}
\end{equation}
Since $G$ is a circulation, Eq.~\ref{def_circ} holds, and summing it over all $i\in\bm{P}$ yields 
\begin{equation}
\sum_{i\in\bm{P},j\in\bm{V}} m_{ij}=\sum_{i\in\bm{P},j\in\bm{V}} m_{ji}
\end{equation}
which can be rewritten as
\begin{equation}
\sum_{i\in\bm{P},j\in\bm{P}} m_{ij}+
\sum_{i\in\bm{P},j\in\bm{V}\setminus\bm{P}} m_{ij}=\sum_{i\in\bm{P},j\in\bm{P}} m_{ji}+
\sum_{i\in\bm{P},j\in\bm{V}\setminus\bm{P}} m_{ji}
\end{equation}
and thus
\begin{equation}
\sum_{i\in\bm{P},j\in\bm{V}\setminus\bm{P}} m_{ij}=\sum_{i\in\bm{P},j\in\bm{V}\setminus\bm{P}} m_{ji}
\end{equation}
so that Eq.~\ref{ratDelta} becomes
\begin{equation}
\frac{\delta_-(\bm{P})}{\delta_+(\bm{P})}=\gamma
\label{ratDelta2}
\end{equation}
and thus $P$ performs a random walk with forward bias $\gamma^{-1}$.

$(2) \Rightarrow (3)$ can be proved as for the clique (see section~\ref{ana_clique} and Ref.~\cite{Traulsen10}).

$(3) \Rightarrow (4)$ can be proved using conditional probabilities exactly as in Ref.~\cite{Lieberman05}.

$(4) \Rightarrow (1)$ can also be proved similarly as in Ref.~\cite{Lieberman05}. Specifically, using Eq.~\ref{(4)} for $P=1$ and $P'=2$ gives
\begin{equation}
\Phi(\gamma, G,1,2)=\frac{1}{1+\gamma}\,,
\label{int1}
\end{equation}
but denoting by $v$ the initially mutant deme, we can also write the probability that 2 demes become mutant after any number $k$ of migration events as
\begin{equation}
\Phi(\gamma, G,1,2)=\sum_{k=0}^\infty \left[1-\delta_-(v)-\delta_+(v)\right]^k\delta_+(v)=\frac{\delta_+(v)}{\delta_+(v)+\delta_-(v)}\,,
\label{int2}
\end{equation}
and comparing Eqs.~\ref{int1} and~\ref{int2} shows that for any initially mutant deme $v$,
\begin{equation}
\frac{\delta_-(v)}{\delta_+(v)}=\gamma\,.
\end{equation}
But Eq.~\ref{ratDelta} yields
\begin{equation}
\frac{\delta_-(v)}{\delta_+(v)}=\gamma\,\, \frac{\sum_{ j\in\bm{V}\setminus v} m_{ji}}{\sum_{ j\in\bm{V}\setminus v}  m_{ij}}\,,
\end{equation}
and therefore, for all $v$,
\begin{equation}
\sum_{ j\in\bm{V}\setminus v} m_{ji}=\sum_{ j\in\bm{V}\setminus v}  m_{ij}\,,
\end{equation}
which entails
\begin{equation}
\sum_{ j\in\bm{V}} m_{ji}=\sum_{ j\in\bm{V}}  m_{ij}\,,
\end{equation}
and thus $G$ is a circulation (see Eq.~\ref{def_circ}).

\section{Comparison with the model of Ref.~\cite{Houchmandzadeh11}}
\label{compB}

In Ref.~\cite{Houchmandzadeh11}, a model generalizing that of Ref.~\cite{Lieberman05} to the case where each node of the graph is occupied by a deme with a fixed number of individuals was introduced. In the models of Refs.~\cite{Lieberman05} and~\cite{Houchmandzadeh11}, each elementary event is composed of a death event in one deme and a birth event in another one, thus allowing to maintain constant the population of each deme. Furthermore, the order employed to choose the individual that dies and the one that divides matters for final results, yielding Birth-death and death-Birth models, as in the model introduced in Ref.~\cite{Lieberman05} (see Refs.~\cite{Kaveh15, Hindersin15, Pattni15}). Conversely, in our model, migration, death and birth events are all independent. This is made possible by allowing the population size of each deme to vary. Here, we present the model of Ref.~\cite{Houchmandzadeh11} and compare it to our model.

Let us consider wild-type fitness as reference and set it to 1, and let us denote mutant fitness by $1+s$. Let us denote the total number of individuals in deme $i$ by $N_i$, and the number of mutant individuals in deme $i$ by $n_i$. Migration probabilities $w_{ij}$ from deme $i$ to deme $j$ are defined along each edge $ij$ of the graph. We will denote by $T^+_i(\bm{n})$ the transition probability from $n_i$ to $n_i+1$ and by $T^-_i(\bm{n})$ the transition probability from $n_i$ to $n_i-1$, which both depend on the complete state of the system $\bm{n}=(n_1,n_2,\dots,n_M)$. 

\subsubsection{Birth-death dynamics}  
In Birth-death dynamics (also known as ``biased invasion process''~\cite{Antal06,Houchmandzadeh11}), the $w_{ki}$ satisfy the normalization constraint
\begin{equation} \label{eq: normalization BD}
\sum_{i=1}^D w_{ki} = 1.
\end{equation}
In this dynamics, an individual is chosen for reproduction among all the individuals of the population according to its fitness. Assuming that it belongs to island $k$, its offspring migrates to island $i$ with probability $w_{ki}$, where it replaces an individual chosen uniformly at random among the $N_i$ individuals there. The transition probability $T^+_i(\mathbf{n})$ is given by
\begin{equation} \label{eq:BD T+}
T^+_i(\bm{n}) = \underbrace{\frac{N_i-n_i}{N_i}}_{\text{(2)}} \sum_{k=1}^D \underbrace{w_{ki}\, \frac{n_k(1+s)}{\sum_j N_j + s n_j}}_{\text{(1)}}
\end{equation}
where (1) is the probability for a mutant to reproduce on island $k$ and to migrate to $i$, which is then summed over all the islands $k$, and (2) is the probability that, given that a death event occurs on island $i$ (because an individual in $i$ is being replaced), a wildtype individual dies. Analogously:
\begin{equation} \label{eq:BD T-}
T^-_i(\bm{n}) = \frac{n_i}{N_i} \sum_{k=1}^D w_{ki}\, \frac{N_k-n_k}{\sum_{j} N_j + s  n_j}.
\end{equation}

\subsubsection{Death-birth dynamics} 
In death-Birth dynamics (also known as ``biased voter model''~\cite{Antal06,Houchmandzadeh11}), we assume that $\sum_k w_{ki}=1$. In this dynamics, an individual is chosen uniformly at random in the entire population to die. Assuming that death occurred in island $i$, one may consider that a migration event then occurs from island $k$ to $i$ with probability $w_{ki}$. But one may also assume that migration occurs from $k$ to $i$ with a probability proportional to the product of $w_{ki}$ and the total fitness $N_k + sn_k$ of island $k$. The first choice considers fitness to be relevant only within each island, while the second one takes into account fitness across the islands. We will consider the second one because it allows to recover the usual death-Birth model~\cite{Kaveh15, Hindersin15, Pattni15} when $N_i=1\,\forall i\in\{1,\dots,D\}$. Finally, the reproducing individual on island $i$ is chosen according to its fitness within the island. The transition probability $T^+_i(\bm{n})$ reads
\begin{equation}
T^+_i(\bm{n}) = \underbrace{\frac{N_i-n_i}{\sum_j N_j}}_{\text{(1)}} \, \sum_{k=1}^D \underbrace{ \phantom{\bigg(} \frac{w_{ki}(N_k+s n_k)}{\sum_j w_{ji}(N_j+s n_j)} \phantom{\bigg)}}_{\text{(2)}}\, \underbrace{ \frac{n_k(1+s)}{N_k + sn_k} }_{\text{(3)}} = \frac{N_i-n_i}{\sum_j N_j} \, \frac{\sum_{k} w_{ki}n_k(1+s)}{\sum_j w_{ji}(N_j+s n_j)} 
\end{equation}
where (1) is the probability for a wildtype individual to die on island $i$, while (2) is the probability that, given that a death event occurs on island $i$, a migration event occurs from island $k$ to $i$, which takes into account the total fitness of island $k$.
Finally, (3) is the probability that, given that a reproduction event happens in island $k$, it is a mutant who reproduces.  
Similarly,
\begin{equation}
T^-_i(\bm{n}) = \frac{n_i}{\sum_j N_j} \frac{\sum_k w_{ki}(N_k-n_k)}{\sum_j w_{ji}(N_j+s n_j)}\, .
\end{equation}

\subsubsection{Clique} 
Consider a clique made of $D$ demes of size $N$ (all of identical and composition-independent size), such that $w_{ij}=w$ for all $i\neq j$ and $w_{ii}=w'$ for all $i$. If migrations between different demes are rare enough, one can coarse-grain the process and consider that each deme is either fully mutant or fully wild-type, and the state of the clique can then be fully described by the number $\xi$ of mutant demes, which changes when migrations followed by fixation occur. In the Birth-death model, starting from Eq.~\ref{eq:BD T-} and summing over mutant demes, we obtain
\begin{equation}
T_\xi^+=\frac{\xi(D-\xi)}{D+s\xi}w(1+s)\rho_M\,,
\end{equation}
and similarly
\begin{equation}
T_\xi^-=\frac{\xi(D-\xi)}{D+s\xi}w\rho_W\,,
\end{equation}
which entails that
\begin{equation}
\gamma=\frac{T_\xi^-}{T_\xi^+}=\frac{(1+s)\rho_M}{\rho_W}=\frac{\rho_Mf_M}{\rho_W f_W}\,,
\label{gamma_Bahram}
\end{equation}
and thus the probability that the mutant fixes in the whole population starting from $i$ mutant demes is given by Eq.~\ref{Phi_SI} but with $\gamma$ expressed in Eq.~\ref{gamma_Bahram}. The same result is obtained in the death-Birth case.

\subsubsection{Star} 
We consider the star graph with self-loops (i.e. allowing replacement within a given deme, corresponding to migration from this deme to itself, $w_{ii}>0$). Indeed, the rare migration regime that we study in our model means in the framework of the model of Ref.~\cite{Houchmandzadeh11} that replacement is much more frequent within a deme than across two demes, thus requiring very strong self-loops, i.e. large $w_{ii}$ values. While the star with self-loops was introduced in Ref.~\cite{Adlam15} with one individual per node of the graph, in the spirit of Ref.~\cite{Lieberman05}, here we treat it in the model of Ref.~\cite{Houchmandzadeh11}, where each node contains a deme with fixed size $N$. Following Ref.~\cite{Adlam15}, we introduce two parameters $x$ and $y$ ($0<x,y\leq1$) such that $1-y$ is the weight of the self-loop of the center and $1-x$ is the weight of the self-loops on each leaf. The other weights are chosen in order to respect the symmetry of the star and for $W$ to be right stochastic in the Birth-death case and left stochastic in the death-Birth case. Hence, in the Birth-death model, the matrix of migration probabilities reads:
\begin{equation} \label{BD matrix}
W =
\begin{pmatrix}
1-y & y/(D-1) & y/(D-1) & \cdots & y/(D-1) & y/(D-1)\\
x & 1-x & 0 &\cdots & 0 & 0\\
x & 0 & 1-x &\cdots & 0 & 0\\
\vdots & \vdots & \vdots & \ddots & \vdots & \vdots \\
x & 0 & 0 & \cdots & 1-x & 0\\
x & 0 & 0 & \cdots & 0 & 1-x
\end{pmatrix},
\end{equation}
where nodes are numbered so that the first one is the center of the star and others are leaves. The case $x=y=1$ corresponds to the star without self-loops introduced in Ref.~\cite{Lieberman05}. 

In order to compare our model to the model of Ref.~\cite{Houchmandzadeh11} in the case of the star, we choose their respective parameters so that in each deme, both models have the same value for the ratio of the migration rate $T_{mig}$ leaving the deme to the reproduction rate $T_{rep}$ in the same deme. In our model, the reproduction rate per individual is given by $T_{rep}=f_W(1-N_W/K)$ for each deme, whatever its type (leaf or center) -- in the wild-type case. Still in our model, the migration rate leaving the center (to any leaf) is $m_O(D-1)$ per individual, and that leaving a leaf (to the center, which is the only possibility) is $m_I$ per individual. In the framework of the Birth-death model of Ref.~\cite{Houchmandzadeh11}, the total reproduction probability per individual in a deme (irrespective of where the offspring from this deme migrates) is equal to $\sum_j w_{ij}=1$, for both the center and for a leaf, while the total migration probability per individual leaving the center is $\sum_{j\neq 1} w_{1j}=y$ and the one leaving a leaf is $\sum_{j\neq i} w_{ij}=x$ with $i>1$ (see Eq.~\ref{BD matrix}). Thus, to match our model with the Birth-death model of Ref.~\cite{Houchmandzadeh11}, we have the following two constraints: 
\begin{equation}
y=\frac{m_O(D-1)}{f_W(1-N_W/K)}\,,\label{m1}
\end{equation} 
and 
\begin{equation}
x=\frac{m_I}{f_W(1-N_W/K)}\,.\label{m2}
\end{equation}  

\begin{figure}[htb]
	\centering
	\includegraphics[width=0.9\textwidth]{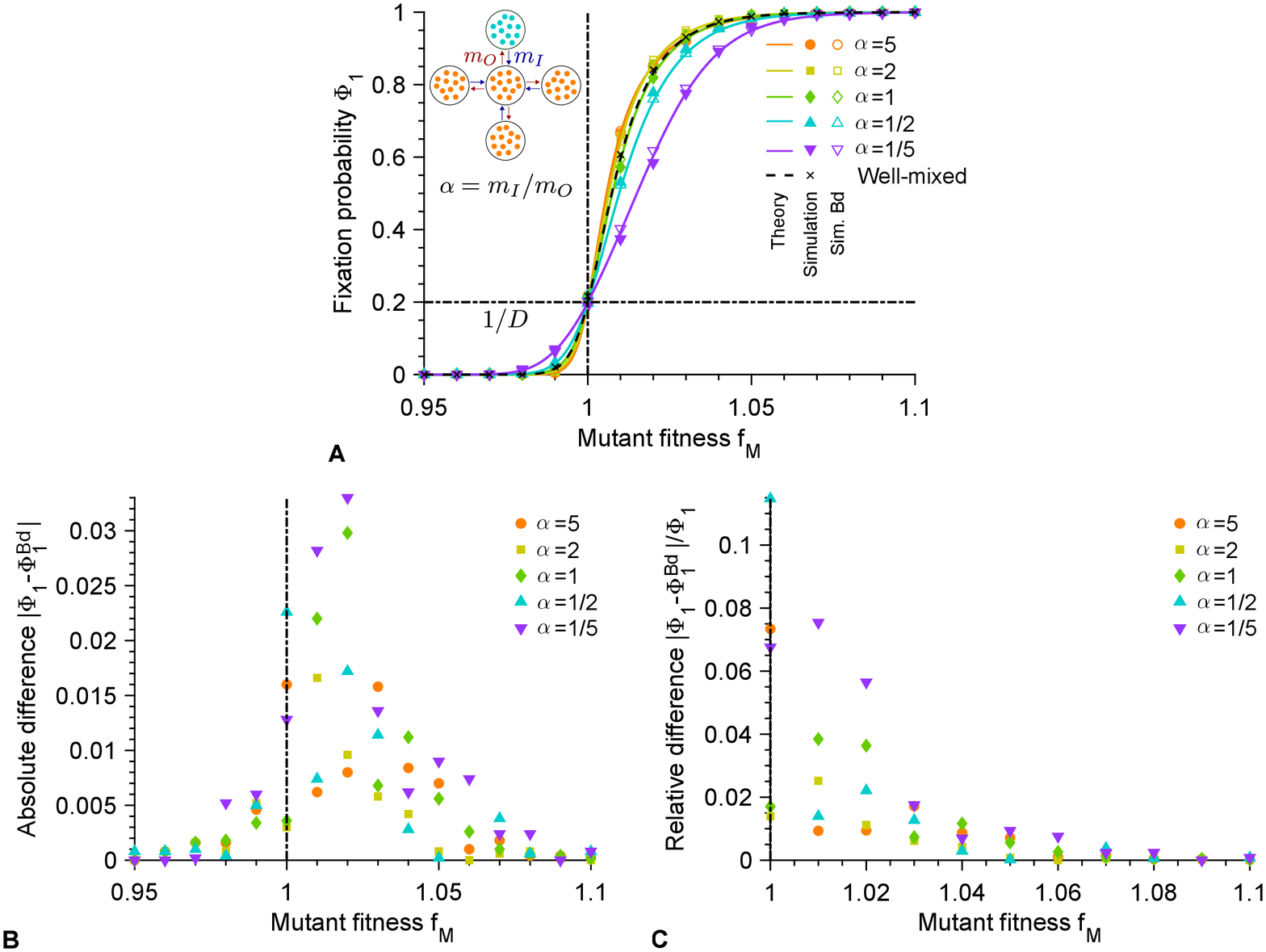}
	\caption{{\bf Comparison between the Birth-death model inspired by Refs.~\cite{Houchmandzadeh11} and~\cite{Adlam15} and our model.} {\bf A:} Fixation probability $\Phi_1$ of mutants in a star graph versus mutant fitness $f_M$, starting with one fully mutant deme chosen uniformly at random, with different migration rate asymmetries $\alpha=m_I/m_O$ in our model and in the matching Birth-death (Bd) model, which satisfies Eqs.~\ref{m1} and~\ref{m2}. Markers are obtained from $2\times 10^3$ stochastic simulation realizations in our model, and in the Bd model of Ref.~\cite{Houchmandzadeh11}. Curves represent analytical predictions for our model in Eqs.~\ref{Phi10},~\ref{Phi01} and~\ref{PhiStar}. {\bf B:} Absolute differences between simulation results obtained with the two models (see panel {\bf A}), as a function of the mutant fitness $f_M$. {\bf C:} Relative differences between simulation results obtained with the two models (see panel {\bf A}), as a function of the mutant fitness $f_M$. Parameter values for our model: $D=5$, $K=100$, $f_W=1$, $g_W=g_M=0.1$, and from top to bottom in the legend of panel {\bf A}, $(m_I, m_O) \times 10^6=(5, 1)$; $(2, 1)$; $(1, 1)$; $(1, 2)$; $(1, 5)$ in simulations, as in Fig.~\ref{Results_star}. Parameter values for the matching Birth-death model: $D=5$, $N=N_W=90$, $f_W=1$, and values of $x$ and $y$ satisfying Eqs.~\ref{m1} and~\ref{m2} for each pair of values of $m_I$ and $m_O$ from our model. Vertical dash-dotted lines indicate the neutral case $f_W=f_M$. }
	\label{comp}%
\end{figure}

Fig. \ref{comp}A shows that once this matching is done, a good agreement is obtained between simulation results for the two models, which yield similar mutant fixation probabilities across various migration asymmetries $\alpha$. Figs. \ref{comp}B and C show small relative and absolute differences, respectively, between the two models. Note that the relative error is high when the probability of fixation of a mutant deme is close to zero, which is the case for deleterious mutations, but then the absolute error is small, which confirms that these models are consistent.


The Birth-death model of Ref.~\cite{Houchmandzadeh11} has the same total reproduction rate in each deme. Once the matching in Eqs.~\ref{m1} and~\ref{m2} is done, it also features the same migration-to-reproduction ratio as in our model. Note however that the death rate is not uniform across demes in this model: in the center it is $\sum_{i} w_{i1}=1-y+(D-1)x$, and in a leaf it is $\sum_{i} w_{ij}=1-x+y/(D-1)$ with $j>1$. This stands in contrast with our model, and to resolve this discrepancy, we would need to impose that $y=(D-1)x$. In that case, the matrix of migration probabilities in Eq.~\ref{BD matrix} becomes doubly stochastic and the star becomes a circulation, and thus it has the same fixation probability as the clique in the model of Ref.~\cite{Houchmandzadeh11} (see above). Consistently, Eqs.~\ref{m1} and~\ref{m2} then entail $\alpha=1$. This shows that the matching between models is not perfect for other values of $\alpha$, because the model of Ref.~\cite{Houchmandzadeh11} is more constrained than our model, as it imposes constant deme size.

In the death-Birth model, the matrix of migration probabilities is the transpose of that given in Eq.~\ref{BD matrix}. Hence, the total reproduction rates for a leaf and the center are $\sum_{i} w_{ij}=y/(D-1)+1-x$ with $j>1$ and $\sum_{i} w_{i1}=1-y+(D-1)x$, respectively, while the total migration rates per individual from a leaf and from the center are $y/(D-1)$ and $(D-1)x$, respectively. Thus, to match our model with the death-Birth model, we have the following two constraints: 
\begin{equation}
xf_W\left(1-\frac{N_W}{K}\right)=m_O(1-y+(D-1)x)\,,\label{m1b}
\end{equation}
and 
\begin{equation}
\frac{y}{D-1}f_W\left(1-\frac{N_W}{K}\right)=m_I\left(\frac{y}{D-1}+1-x\right)\,.\label{m2b}
\end{equation}

\begin{figure}[hbt]
	\centering
	\includegraphics[width=.9\textwidth]{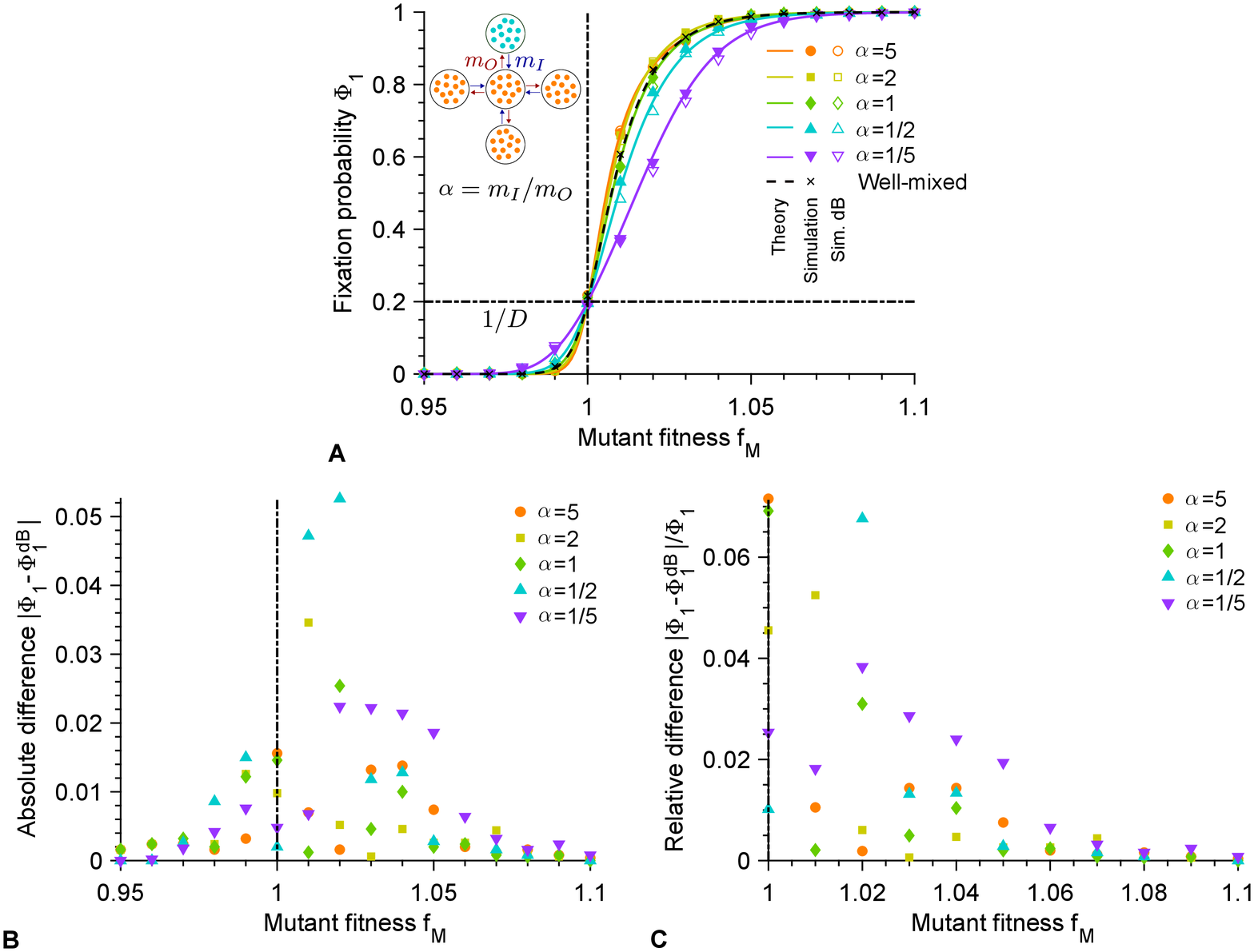}
	\caption{{\bf Comparison between the death-Birth model inspired by Refs.~\cite{Houchmandzadeh11} and~\cite{Adlam15} and our model.} {\bf A:} Fixation probability $\Phi_1$ of mutants in a star graph versus mutant fitness $f_M$, starting with one fully mutant deme chosen uniformly at random, with different migration rate asymmetries $\alpha=m_I/m_O$ in our model and in the matching death-Birth (dB) model, which satisfies Eqs.~\ref{m1b} and~\ref{m2b}. Markers are obtained from $2\times 10^3$ stochastic simulation realizations in our model, and in the dB model of Ref.~\cite{Houchmandzadeh11}. Curves represent analytical predictions for our model in Eqs.~\ref{Phi10},~\ref{Phi01} and~\ref{PhiStar}. {\bf B:} Absolute differences between simulation results obtained with the two models (see panel {\bf A}), as a function of the mutant fitness $f_M$. {\bf C:} Relative differences between simulation results obtained with the two models (see panel {\bf A}), as a function of the mutant fitness $f_M$. Parameter values for our model: $D=5$, $K=100$, $f_W=1$, $g_W=g_M=0.1$, and from top to bottom in the legend of panel {\bf A}, $(m_I, m_O) \times 10^6=(5, 1)$; $(2, 1)$; $(1, 1)$; $(1, 2)$; $(1, 5)$ in simulations, as in Fig.~\ref{Results_star}. Parameter values for the matching death-Birth model: $D=5$, $N=N_W=90$, $f_W=1$, and values of $x$ and $y$ satisfying Eqs.~\ref{m1b} and~\ref{m2b} for each pair of values of $m_I$ and $m_O$ from our model. Vertical dash-dotted lines indicate the neutral case $f_W=f_M$. }
	\label{comp2}%
\end{figure}

In this case too, Fig. \ref{comp2} shows that once this matching is done, a good agreement is obtained between simulation results for the two models. 


The death-Birth model of Ref.~\cite{Houchmandzadeh11} has the same total death rate in each deme. Once the matching in Eqs.~\ref{m1} and~\ref{m2} is done, it also features the same migration-to-reproduction ratio as in our model. Note however that the birth rate is not uniform across demes in this model even in the absence of fitness differences (see above). Here too, to resolve this discrepancy with our model, we would need to impose that $y=(D-1)x$, with the same consequences as in the Birth-death model -- note that Birth-death and death-Birth models then yield the same result. Again, this shows that the matching between models is not perfect for other values of $\alpha$, because the model of Ref.~\cite{Houchmandzadeh11} is more constrained than our model, as it imposes constant deme size.


\section{Extension to different deme sizes: the doublet}\label{ana_two_pop}

\subsection{Main results}

Our model allows us to consider structures involving demes with different sizes. In this case, we consider an initial mutant placed randomly with a probability proportional to deme size, which is realistic for mutations occurring upon division or with a constant rate per individual (note that this corresponds to both uniform and temperature initial conditions in the language of models with a single individual per node~\cite{Adlam15}, which coincide in our model).

As a simple example, consider a doublet comprising a small deme with carrying capacity $K_S$ and a larger deme with carrying capacity $K_L>K_S$ (see Fig. \ref{Struct}D). Individuals can migrate from the large (resp. small) deme to the small (resp. large) deme with a rate per individual $m_S$ (resp. $m_L$). For structured populations involving demes with identical sizes, we considered the fixation probability $\Phi_1$ starting from one fully mutant deme, which yields  that of one mutant individual when multiplied by $\rho_M$. Here, we consider the fixation probability of one single mutant in the structure divided by that in the small deme. If we define $D$ such that $K_L=(D-1)K_S$ and if we choose the notation $K_S=K$, then this quantity $\Phi_1^\textrm{doublet}$ is analogous to our usual $\Phi_1$ (if $D$ is an integer), thus facilitating comparisons. $\Phi_1^\textrm{doublet}$ is expressed analytically below. 

\begin{figure}[htb]
	\centering
	\includegraphics[width=.5\textwidth]{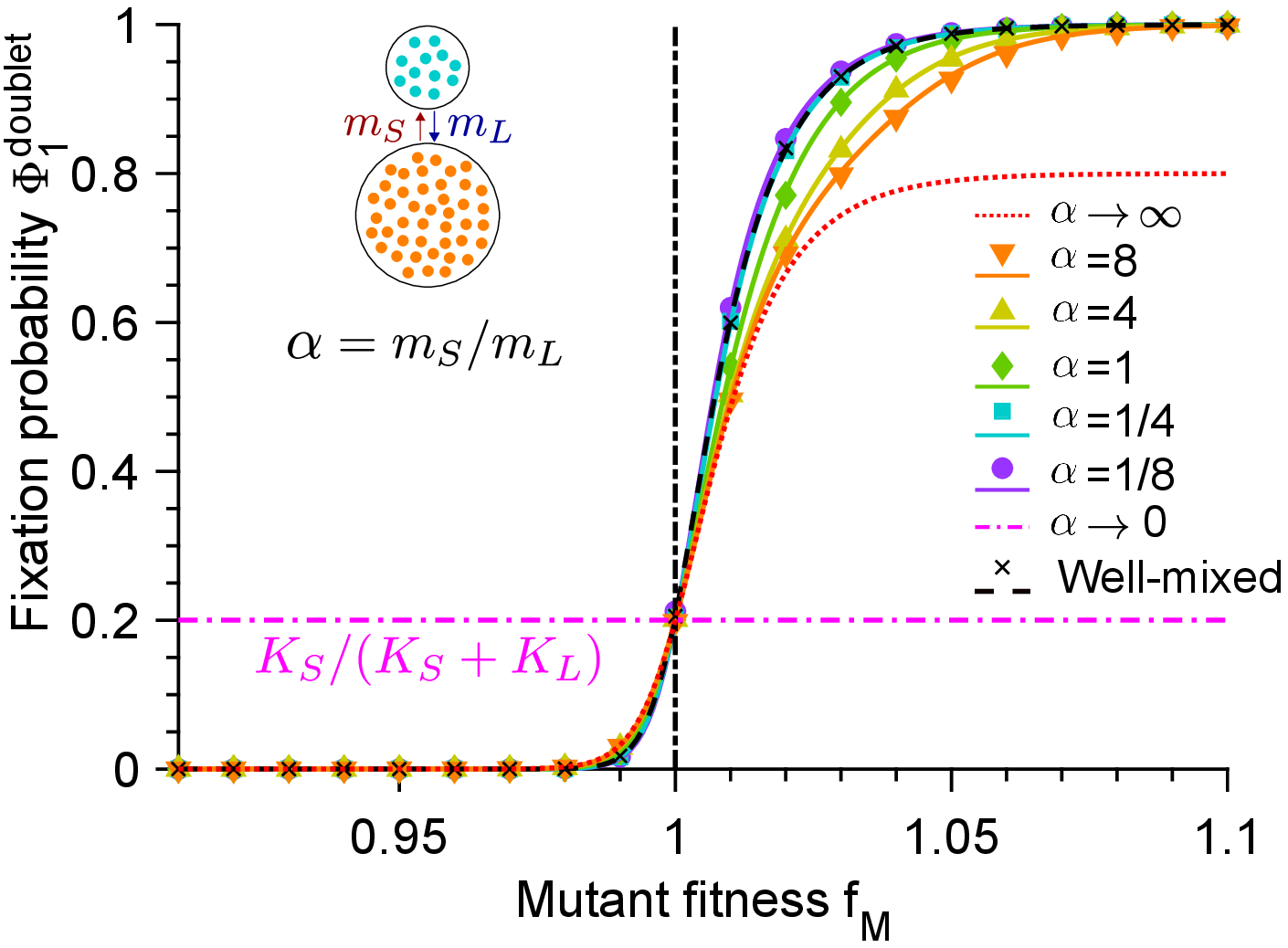}
	\vspace{0.2cm}
	\caption{{\bf Fixation probability for the doublet.} Fixation probability $\Phi_1^\textrm{doublet}$ of mutants in a doublet versus mutant fitness $f_M$, starting with one fully mutant deme chosen proportionally to deme size, with different migration asymmetries $\alpha=m_S/m_L$. Data for the well-mixed population is shown as reference, with same total population size and initial number of mutants. Markers are computed over $2\times 10^3$ stochastic simulation realizations. Curves represent analytical predictions in Eqs.~\ref{PhiStwopop}, \ref{PhiLtwopop} and \ref{Phi1twopop}. Vertical dash-dotted lines indicate the neutral case $f_W=f_M$. Parameter values: $K_S=100$, $K_L=400$ (hence $K_L=(D-1)K_S$ with $K_S=K=100$ and $D=5$), $f_W=1$, $g_W=g_M=0.1$. From top to bottom, $(m_S, m_L) \times 10^6=(8, 1)$; $(4, 1)$; $(1, 1)$; $(1, 4)$; $(1, 8)$ in simulations.}
	\label{Results_two_pop}%
\end{figure}

\begin{figure}[htb]
	\centering
	\includegraphics[width=\textwidth]{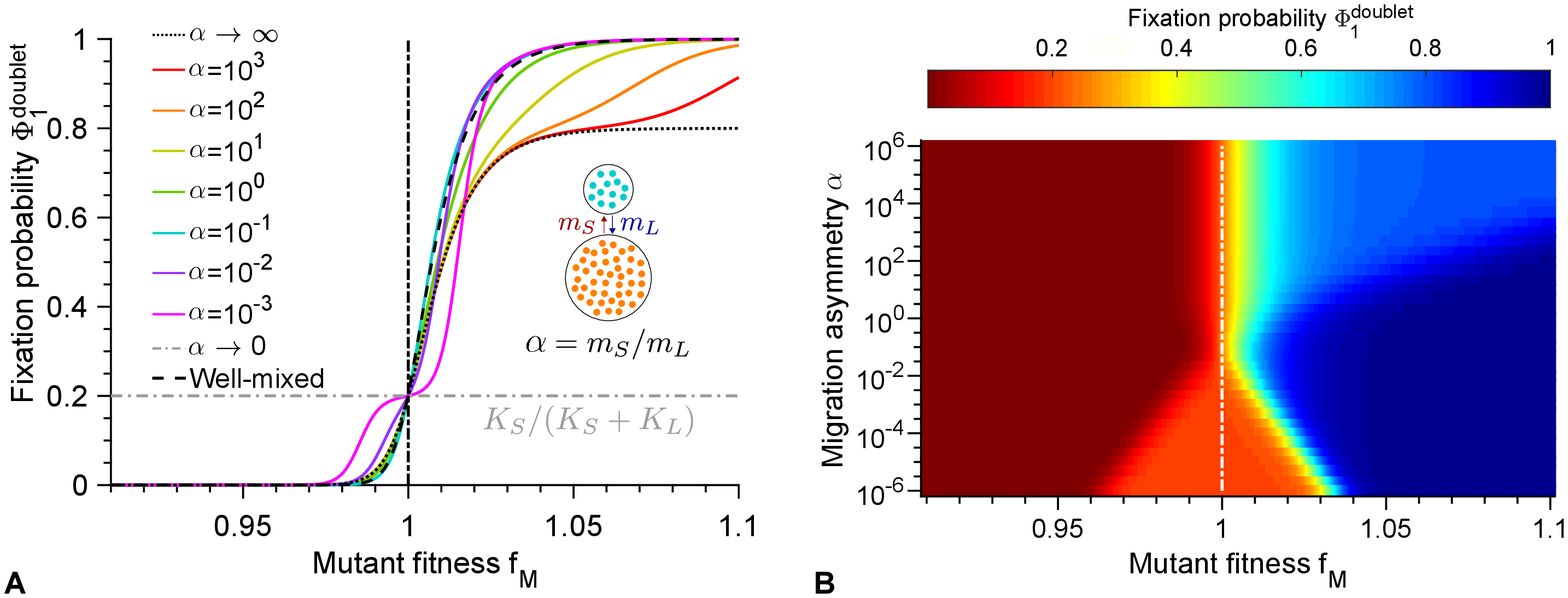}
	\caption{{\bf Fixation probability for the doublet.} {\textbf{A}:} Fixation probability $\Phi_1^\textrm{doublet}$ of mutants in a doublet versus mutant fitness $f_M$, starting with one fully mutant deme chosen proportionally, with different migration rate asymmetries $\alpha=m_S/m_L$, complementing those shown in Fig.~\ref{Results_two_pop}. Data for the well-mixed population is shown as reference, with same total population size and initial number of mutants. Curves represent analytical predictions in Eqs.~\ref{PhiStwopop}, \ref{PhiLtwopop} and \ref{Phi1twopop}. {\textbf{B}:} Heatmap of the same fixation probability versus mutant fitness $f_M$ and migration rate asymmetry $\alpha=m_S/m_L$. Parameter values in both panels: $K_S=100$, $K_L=400$ (hence $K_L=(D-1)K_S$ with $K_S=K=100$ and $D=5$), $f_W=1$, $g_W=g_M=0.1$. Vertical dash-dotted lines represent the neutral case $f_W=f_M$.}
	\label{Results_two_pop_more}%
\end{figure}

Fig. \ref{Results_two_pop} shows $\Phi_1^\textrm{doublet}$ for different migration asymmetries $\alpha=m_S/m_L$, with excellent agreement between our analytical predictions and our simulation results (see also Fig. \ref{Results_two_pop_more} for additional $\alpha$ values and a heatmap). Furthermore, Fig. \ref{Results_two_pop} shows that the fixation probability is  very close to the well-mixed case when $\alpha=1/(D-1)$. This corresponds to $m_SK_L=m_LK_S$, i.e. to equal migration flows from small to large deme and reciprocally. We also observe that the doublet behaves as a suppressor of selection for $\alpha>1/(D-1)$, and has weak amplifying properties for  $\alpha<1/(D-1)$, which do not survive in the limit $\alpha\to 0$. In the Appendix, Section~\ref{ana_two_pop}, we show that in the regime of moderate mutational effects, the doublet is an amplifier of selection with respect to the clique for $1/(D-1)^2<\alpha<1/(D-1)$, and a suppressor of selection for $\alpha>1/(D-1)$. This generalizes the result of Ref.~\cite{Lieberman05} that small upstream populations with large downstream populations, corresponding here to $\alpha\to 0$, yield suppressors. Furthermore, this confirms the importance of migration asymmetry in the impact a population structure has on selection. 

\newpage

\subsection{Fixation probability}

\subsubsection{General expression}

In order to calculate the fixation probability of the mutant type in the doublet, let us first consider the case where the small deme, whose carrying capacity is denoted by $K_S$, is fully mutant, while the large deme, whose carrying capacity is denoted by $K_L$, is fully wild-type. Recall that the migration rate per individual from the small deme to the large one is $m_S$, and that from the large to the small deme by $m_L$. We start from exactly one fully mutant deme. If an $M$ individual migrates from the small deme to the large deme and fixes, then the mutant type fixes in the whole population. The probability that this occurs upon a given migration event reads \begin{equation}
T_S^+=\frac{m_LN_M^S}{m_LN_M^S+m_SN_W^L}\rho_M^L\,,
\end{equation} 
where $N_M^S=K_S(1-g_M/f_M)$ (respectively $N_W^L=K_L(1-g_W/f_W)$) is the equilibrium size of the small mutant deme (respectively of the large wild-type deme) and $\rho_M^L$ is the fixation probability of a mutant in the large wild-type deme, given by Eq.~\ref{rhom} with $N_W^L$ instead of $N_W$. Similarly, if a $W$ individual migrates to the small deme and fixes, then the wild-type fixes in the whole population. The probability that this occurs upon a given migration event reads
\begin{equation}
T_S^-=\frac{m_SN_W^L}{m_LN_M^S+m_SN_W^L}\rho_W^S\,,
\end{equation}  
where $\rho_W^S$ is the fixation probability of a wild-type individual in the small mutant deme, given by Eq.~\ref{rhow} with $N_M^S$ instead of $N_M$. Then, the fixation probability of the mutant type, starting from a small mutant deme and a large wild-type deme, reads 
\begin{equation}
\Phi_{1,S}^\textrm{doublet}=\frac{T_S^+}{T_S^++T_S^-}=\frac{m_LN_M^S\rho_M^L}{m_LN_M^S\rho_M^L+m_SN_W^L\rho_W^S}=\frac{1}{1+\alpha\gamma_S }\,, 
\label{PhiStwopop}
\end{equation}  
where $\alpha=m_S/m_L$ and $\gamma_S=N_W^L\rho_W^S/(N_M^S\rho_M^L)$.

Similarly, in the case where the structured population starts from a large mutant deme, while the small deme is wild-type, we get the fixation probability 
\begin{equation}
\Phi_{1,L}^\textrm{doublet}=\frac{m_SN_M^L\rho_M^S}{m_SN_M^L\rho_M^S+m_LN_W^S\rho_W^L}=\frac{1}{1+\gamma_L/\alpha}\,,
\label{PhiLtwopop}
\end{equation}
where $\gamma_L=N_W^S\rho_W^L/(N_M^L\rho_M^S)$.

Next, consider the case where one mutant individual starts in a deme with a probability proportional to the size of the deme, which corresponds to the realistic case of mutations happening randomly upon division. The fixation probability of such a single mutant reads:
\begin{equation}
\rho_M^\textrm{doublet}=\frac{K_S}{K_S+K_L}\rho_{M}^S\Phi_{1,S}^\textrm{doublet}+\frac{K_L}{K_S+K_L}\rho_{M}^L\Phi_{1,L}^\textrm{doublet}\,.
\end{equation}
In the rest of this work, which focuses on structured populations made of demes of identical sizes, we consider the fixation probability $\Phi_1$ starting from one fully mutant deme, which  then needs to be multiplied by $\rho_M$ to obtain that of one mutant individual. Here, we will consider the analogous quantity
\begin{equation}
\Phi_1^\textrm{doublet}=\frac{\rho_M^\textrm{doublet}}{\rho_M^S}=\frac{K_S}{K_S+K_L}\Phi_{1,S}^\textrm{doublet}+\frac{K_L}{K_S+K_L}\frac{\rho_{M}^L}{\rho_{M}^S}\Phi_{1,L}^\textrm{doublet}\,.
\label{Phi1twopop}
\end{equation}
In the particular case where $K_L=(D-1)K_S$, so that the total carrying capacity of the subdivided population is $DK_S$, denoting $K_S$ by $K$, considering $\Phi_1^\textrm{doublet}$ allows for a direct comparison to $\Phi_1$ in the other structures considered here, comprising $D$ demes of carrying capacity $K$.

\subsubsection{Expansion for relatively small mutational effects} 
For the sake of simplicity, here we assume that $K_L=(D-1)K_S$, so that the total carrying capacity of the subdivided population is $DK_S$, and we further denote $K_S$ by $K$. Consider the regime where $\epsilon\ll 1$ and $N_W|\epsilon|\gg 1$ but $N_W\epsilon^2\ll 1$. Then, if $\epsilon>0$, Eqs.~\ref{PhiStwopop}, \ref{PhiLtwopop} and \ref{Phi1twopop} yield
\begin{equation}
\Phi_1^\textrm{doublet}=1-(\alpha+1)\frac{D-1}{D}e^{-N_W\epsilon}\left[1+O(\epsilon)+O(N_W\epsilon^2)\right]\,,
\label{phiDbl1Pos}
\end{equation}
which gives, employing Eq.~\ref{phiClique1Pos},
\begin{equation}
\frac{\Phi_1^\mathrm{doublet}}{\Phi_1^\mathrm{clique}}=1+\frac{1-\alpha\left(D-1\right)}{D}e^{-N_W\epsilon}\left[1+O(\epsilon)+O(N_W\epsilon^2)\right]\,.
\label{ratioDbl1Pos}
\end{equation}
Thus, in this case, assuming $D>1$, we have $\Phi_1^\mathrm{doublet}>\Phi_1^\mathrm{clique}$ if $\alpha<1/(D-1)$, whereas $\Phi_1^\mathrm{doublet}<\Phi_1^\mathrm{clique}$ if $\alpha>1/(D-1)$. Now if $\epsilon<0$, Eqs. \ref{PhiStwopop}, \ref{PhiLtwopop} and \ref{Phi1twopop} yield
\begin{equation}
\Phi_1^\textrm{doublet}=\frac{1+\alpha^2\left(D-1\right)^3}{\alpha D\left(D-1\right)}e^{N_W(D-1)\epsilon}\left[1+O(\epsilon)+O(N_W\epsilon^2)\right]\,,
\label{phiDbl1Neg}
\end{equation}
which gives, employing Eq.~\ref{phiClique1Neg},
\begin{equation}
\frac{\Phi_1^\mathrm{doublet}}{\Phi_1^\mathrm{clique}}=\frac{1+\alpha^2\left(D-1\right)^3}{\alpha D\left(D-1\right)}\left[1+O(\epsilon)+O(N_W\epsilon^2)\right]\,.
\label{ratioDbl1Neg}
\end{equation}
Then, assuming $D>2$, studying the function $G:\alpha\mapsto\left[1+\alpha^2(D-1)^3\right]/\left[\alpha D\left(D-1\right)\right]$ demonstrates that $\Phi_1^\mathrm{doublet}>\Phi_1^\mathrm{clique}$ if $\alpha<1/(D-1)^2$ or $\alpha>1/(D-1)$, whereas $\Phi_1^\mathrm{doublet}<\Phi_1^\mathrm{clique}$ if $1/(D-1)^2<\alpha<1/(D-1)$. Therefore, in the regime where $\epsilon\ll 1$ and $N_W|\epsilon|\gg 1$ but $N_W\epsilon^2\ll 1$, the doublet is an amplifier of selection with respect to the clique for $1/(D-1)^2<\alpha<1/(D-1)$, and a suppressor of selection for $\alpha>1/(D-1)$. Finally, for $\alpha<1/(D-1)^2$, it behaves as a suppressor for $\epsilon<0$ and as an amplifier for $\epsilon>0$.

\subsubsection{Expansion for extremely asymmetric migrations} 
If $\alpha\rightarrow 0$, then Eqs.~\ref{PhiStwopop}, \ref{PhiLtwopop} and \ref{Phi1twopop} yield
\begin{equation}
\Phi_1^\mathrm{doublet}\approx \frac{1}{D}\,,
\end{equation}
while if $\alpha\rightarrow\infty$, they give
\begin{equation}
\Phi_1^\mathrm{doublet}\approx \frac{D-1}{D}\frac{\rho_M^L}{\rho_M^S}\,.
\end{equation}

\section{Constant deme size approximation}\label{EquilibriumSize}

In our model, we consider that the number of individuals in each deme is not fixed, but there is a carrying capacity $K$ per deme. In a deterministic description, valid for large populations, if there is only one type of individuals, the number $N$ of individuals at time $t$ follows the ordinary differential equation: 
\begin{equation}
\frac{\mbox{d}N}{\mbox{d}t}=\left[f\left(1-\frac{N}{K}\right)-g\right]N\mbox{ },
\label{dNdt}
\end{equation}
where $f$ represents fitness, $g$ death rate and $K$ carrying capacity. If $f>g$, Eq. \ref{dNdt} yields a nonzero steady-state population size, namely $K(1-g/f)$. In a stochastic description, a finite-size microbial population with a logistic growth rate and a constant death rate fluctuates around the deterministic steady-state average population size $K(1-g/f)$ after a transient time depending on initial conditions and before eventually going extinct (after a very long time if it carrying capacity is not small) \cite{Vogels75, Ovaskainen10}. Therefore, in our analytical studies, we often employ the steady-state population sizes of wild-type and mutant demes, denoted by $N_W$ and $N_M$ respectively:
\begin{equation}
N_W=K(1-g_W/f_W)\mbox{ },
\label{NA}
\end{equation}
and
\begin{equation}
N_M=K(1-g_M/f_M)\mbox{ }.
\label{NB}
\end{equation}

\begin{figure}[h!]
	\begin{center}
		\includegraphics[width=\textwidth]{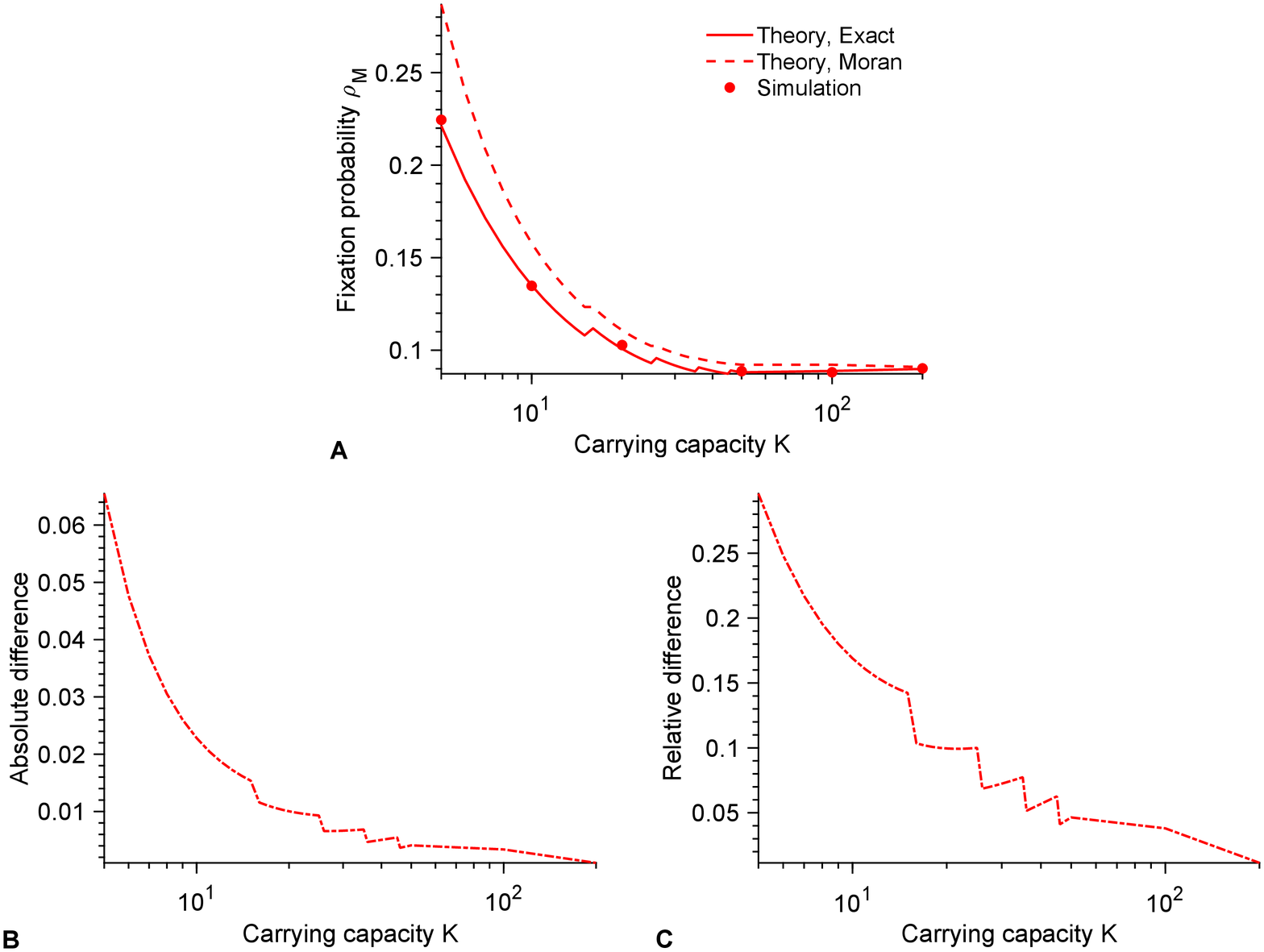}
		\caption{{\bf Constant deme size approximation.} Fixation probability $\rho_M$ of a mutant in a population of wild-type individuals with carrying capacity $K$ and steady-state size $N_W=K(1-g_W/f_W)$. Markers: averages over $10^4$ stochastic simulations. Solid line: numerical resolution of the Master equation, see Eq. 3 in Ref.~\cite{Parsons07}. Dashed line: constant-size approximation employed in this work, in the framework of the Moran process~\cite{Moran58,Ewens79} (see Eq. \ref{rhom}).  Parameter values: $f_W=1$, $f_M=1.1$, $g_W=g_M=0.1$. Absolute and relative differences between the dashed and solid lines of panel {\bf A} are shown in panels {\bf B} and {\bf C}. Discontinuities arise from the need to set the constant population size in the Moran to an integer value, while the steady-state size is not necessarily an integer -- this is done by truncation.} 
		\label{phiAandphiB}
	\end{center}
\end{figure}

Furthermore, for simplicity, we approximate fixation probabilities in each deme by their values computed at fixed population size within the Moran process~\cite{Moran58,Ewens79}. The fixation probability of a single mutant (resp. wild-type) in a wild-type (resp. mutant) deme of steady-state size $N_W$ (resp. $N_M$) is then given by Eq. \ref{rhom} (resp. Eq.~\ref{rhow}). This approximation is expected to be reasonable for large enough steady-state deme sizes. This is confirmed by Fig. \ref{phiAandphiB}, where the constant-size approximation from Eq. \ref{rhom} is compared to results from stochastic simulations of the evolutionary dynamics of a mutant in a population of $W$ individuals with variable population size, and to a numerical resolution of the Master equation for variable population size, based on Ref.~\cite{Parsons07}. In the cases with variable population size, we use a carrying capacity $K$ and a steady-state size $N_W=K(1-g_W/f_W)$, as in the rest of our work.

\clearpage

\section{Simulation methods}\label{Sim}

Implementations of our simulations in the C programming language are freely available at\\ \texttt{https://doi.org/10.5281/zenodo.5126699}.

Our numerical simulations are performed using a Gillespie algorithm that is exact and does not involve any artificial discretization of time \cite{Gillespie76, Gillespie77}. We focus on the regime where deme sizes $N_i$ fluctuate weakly around their deterministic steady-state values, namely $N_i=K_i(1-g_a/f_a)$ if all microbes in deme $i$ are of type $a$.  Thus, we start our simulations at these sizes, and we consider $K_i$ large enough for stochastic extinctions not to occur within the timescales studied. In most cases, we start our simulations with one fully mutant deme, while all others are fully wild-type, because this describes the second step in the fixation of a mutant (after it has fixed in a deme) in the rare migration regime. Note however that our stochastic simulations are valid beyond the rare migration regime and allow us to test the validity of this assumption and to go beyond this regime. We consider a structured population of $D$ demes labeled $i=1,2,...,D$, and denote by $N_{W,i}$ and $N_{M,i}$ the respective numbers of $W$ and $M$ individuals in deme $i$. \\

The elementary events that can happen are reproduction, death and migration of an individual of either type:
\begin{itemize}
	\item $W_i \overset{k_{W,i}^+}{\rightarrow} 2W_i$: Reproduction of a wild-type microbe in deme $i$ with rate $k_{W,i}^{+}=f_W[1-(N_{W,i}+N_{M,i})/K]$. 
	\item $W_i \overset{k_{W,i}^-}{\rightarrow} \emptyset$: Death of a wild-type microbe in deme $i$ with rate $k_{W,i}^{-}=g_W$.
	\item $M_i \overset{k_{M,i}^+}{\rightarrow} 2M_i$: Reproduction of a mutant microbe in deme $i$ with rate $k_{M,i}^{+}=f_M[1-(N_{W,i}+N_{M,i})/K]$. 
	\item $M_i \overset{k_{M,i}^-}{\rightarrow} \emptyset$: Death of a mutant microbe in deme $i$ with rate $k_{M,i}^{-}=g_M$. Note that we take $g_M=g_W$ throughout.
	\item $W_i  \overset{m_{ij}}{\rightarrow}  W_j$: Migration of a wild-type microbe from deme $i$ to deme $j$ with rate $m_{ij}$.
	\item $M_i  \overset{m_{ij}}{\rightarrow} M_j$: Migration of a mutant microbe from deme $i$ to deme $j$ with rate $m_{ij}$.
\end{itemize}
The total rate of events is given by $k_{tot}=\sum_{i=1}^D\left(k_{W,i}^++k_{W,i}^-\right)N_{W,i}+\left(k_{M,i}^++k_{M,i}^-\right)N_{M,i}+\sum_{i,j=1}^D m_{ij}\left(N_{W,i}+N_{M,i}\right)$.\\

Simulation steps are as follows:
\begin{enumerate}
	\item Initialization: All of the $D$ demes start from either $N_W=K(1-g_W/f_W)$ wild-type microbes or $N_M=K(1-g_M/f_M)$ mutant microbes, at time $t=0$.
	\item Monte Carlo step: Time $t$ is incremented by $\Delta t$, sampled from an exponential distribution with mean $1/k_{tot}$. The next event to occur is chosen proportionally to its probability $k/k_{tot}$, where $k$ is its rate, and is executed.
	\item We go back to Step 2 unless only one type of individuals, either $W$ or $M$, remains in the population, which corresponds to fixation of one type. Simulation is ended when fixation occurs.
\end{enumerate}

\newpage

\end{document}